\documentclass[12pt, a4paper]{article}
 
\usepackage{amsmath, amssymb, graphicx}
\usepackage{tikz}
\usetikzlibrary{arrows.meta, positioning, shapes.geometric}
\usepackage[linesnumbered,ruled,vlined]{algorithm2e}
\usepackage{xcolor}
\usepackage{listings}
\usepackage{courier}
\usepackage{fancyvrb}
\usepackage{caption}
\usepackage[margin=1in]{geometry}
\usepackage{natbib}
\usepackage{mathptmx}
\usepackage{hyperref} 
\usepackage{booktabs}

\lstdefinestyle{pythonstyle}{
    language=Python,
    basicstyle=\ttfamily\small,
    keywordstyle=\color{blue},
    commentstyle=\color{gray},
    stringstyle=\color{red!70!black},
    showstringspaces=false,
    breaklines=true,
    frame=single,
    numbers=left,
    numberstyle=\tiny\color{gray},
    captionpos=b,
    tabsize=4
}

\title{
Integrating Multi-Armed Bandit, Active Learning, and Distributed Computing for Scalable Optimization}
\author{ Foo Hui-Mean and Yuan-chin Ivan Chang\\ Institute of Statistical Science\\ Academia Sinica, Taiwan}

\begin{document}
\maketitle

\begin{abstract}
Modern optimization problems in scientific and engineering domains often rely on expensive black-box evaluations, such as those arising in physical simulations or deep learning pipelines, where gradient information is unavailable or unreliable. In these settings, conventional optimization methods quickly become impractical due to prohibitive computational costs and poor scalability.
We propose ALMAB-DC, a unified and modular framework for scalable black-box optimization that integrates active learning, multi-armed bandits, and distributed computing, with optional GPU acceleration. The framework leverages surrogate modeling and information-theoretic acquisition functions to guide informative sample selection, while bandit-based controllers dynamically allocate computational resources across candidate evaluations in a statistically principled manner. These decisions are executed asynchronously within a distributed multi-agent system, enabling high-throughput parallel evaluation. We establish theoretical regret bounds for both UCB-based and Thompson-sampling-based variants and develop a scalability analysis grounded in Amdahl's and Gustafson's laws. Empirical results across synthetic benchmarks, reinforcement learning tasks, and scientific simulation problems demonstrate that ALMAB-DC consistently outperforms state-of-the-art black-box optimizers. By design, ALMAB-DC is modular, uncertainty-aware, and extensible, making it particularly well suited for high-dimensional, resource-intensive optimization challenges.
\end{abstract}

{
\section{Introduction}

Optimization plays a critical role in scientific computing problems in engineering design,  machine learning, and others. Many modern applications involve black-box functions that are costly to evaluate, lack closed-form gradients, and often produce noisy or simulation-based outputs. Typical examples include hyper-parameter tuning in deep neural networks reinforcement learning, and computational fluid dynamics simulations, where traditional gradient-based or exhaustive search methods become impractical due to their high computational demands  \citep{krizhevsky2009learning, kennedy2001bayesian}.

For addressing these challenges, we introduce \textbf{ALMAB-DC}, a unified framework that combines principles from active learning, multi-armed bandit algorithms, and distributed computing, and treats optimization as a sequential decision-making process under uncertainty, guided by statistical surrogates that prioritize informative queries and balance exploration with exploitation.  ALMAB-DC is designed to efficiently navigate high-dimensional spaces under budget constraints by combining adaptive sampling with scalable infrastructure.  This framework contributes a general-purpose optimization strategy that bridges statistical design, online decision theory, and large-scale parallelism, and supports various acquisition mechanisms, integrates multi-agent execution, with optional GPU acceleration for computational efficiency. Its generality allows deployment across a wide range of applications, including hyper-parameter optimization, simulation-based scientific discovery, and resource-constrained experimental design.

%
%

Several recent works have explored partial intersections of the components considered in this study, yet few offer a unified perspective. \cite{auer2022distributed} investigated distributed online learning in multi-armed bandits, focusing on delayed feedback, but without incorporating active learning elements. \cite{bachman2017budgeted} proposed a hypothesis-space-based approach to budgeted active learning, though they did not address scalability or integrate bandit frameworks. \cite{landgren2016distributed} examined decentralized exploration in multi-agent bandit settings, while \cite{li2019distributedAL} focused on active learning in distributed environments, omitting any bandit-based optimization strategy. \cite{liu2021active} introduced active reward mechanisms to improve reinforcement learning efficiency, effectively combining AL with reward shaping. Wang and Liu~\cite{wang2020bandits} presented a unified view of active learning and bandits but did not explore issues of scalability or distributed implementation. \cite{yu2022federated} examined federated active learning through peer selection, and \cite{li2018hyperband} applied bandit algorithms to distributed hyperparameter optimization.

In contrast, the proposed ALMAB-DC framework integrates all three paradigms—active learning, multi-armed bandits, and distributed computing—into a single, scalable optimization architecture. It introduces a principled decomposition of regret across both the active learning and bandit dimensions and demonstrates effective scaling via distributed agents. This level of conceptual integration and theoretical rigor is not evident in prior works. The potential value of ALMAB-DC lies in its generality and extensibility: it can be adapted to a variety of domains, including simulation-based optimization, hyper-parameter tuning, and autonomous scientific discovery, enabling substantial improvements in resource-efficient, large-scale decision-making. 

The remainder of this paper details the system architecture, sampling strategies, and coordination mechanisms that make up the ALMAB-DC pipeline, along with comparative evaluations and discussions of related work.

\section{Methodology}


Instead of using the batch-based approach, ALMAB-DC treats the problem as a sequential decision-making process under uncertainty for large-scale black-box optimization. Each decision—configuration to evaluate—is guided by surrogate models that quantify uncertainty and expected utility, informed by statistical experimental design.
Unlike traditional batch-based optimization methods, ALMAB-DC uses an iterative, data-adaptive approach, and selects the next most informative input,  at each step, to evaluate using acquisition functions such as entropy, expected improvement, mutual information, or other statistical information metrics. These strategies align with the principles of (Bayesian) active learning and are used to minimize sample complexity while maintaining high-quality solutions.
To balance exploration and exploitation efficiently, the framework embeds multi-armed bandit (MAB) scheduling strategies. The decision of which candidate configuration to evaluate is treated as a bandit arm selection problem, enabling principled regret minimization through algorithms such as UCB and Thompson Sampling.
%
%
The system distributes evaluation tasks across a network of parallel agents and supports, as an option, GPU-accelerated computation for operations such as surrogate modeling, posterior updates, and acquisition function optimization. These components operate asynchronously to improve throughput, particularly in resource-constrained or heterogeneous environments. We begin with an overview of ALMAB-DC  below, and a detailed description of the full system architecture and each module follows.

\subsection{ALMAB-DC Framework}
ALMAB-DC is a modular framework that integrates Active Learning (AL), Multi-Armed Bandits (MAB), and Distributed Computing (DC) to efficiently solve expensive black-box optimization problems. It uses a Bayesian surrogate model as a probabilistic approximation of the objective function, guiding sample selection while minimizing redundant evaluations.
The process begins with an \textit{Unlabeled Data Pool} of candidate configurations. The \textit{Active Learner} selects the most informative samples using criteria such as BALD or Core-set, forwarding them to the \textit{Bandit Controller}, which balances exploration and exploitation using strategies like UCB or Thompson Sampling. The selected configuration is dispatched to a \textit{Distributed Agent}, which executes tasks asynchronously across compute nodes using frameworks such as Ray or MPI. For computationally intensive workloads, the optional \textit{GPU- 
We state the role of each modular below.
The arrows in Figure~\ref{fig:almabdc} show how updated predictions from the surrogate model flow into the AL and MAB components, whose decisions guide the distributed evaluations. Accelerated Evaluation Module} handles model training or simulation. The resulting performance metrics are then fed into the \textit{Surrogate Model Update} module, refining the Bayesian model and closing the learning loop.
Figure~\ref{fig:ALMAB-DC-Pipeline} illustrates the overall pipeline, showing the six core modules and their interactions. This architecture supports multi-fidelity modeling, hierarchical surrogate learning, and asynchronous execution, making it scalable and adaptable across diverse computing environments. A detailed summary of key features is provided in Table~\ref{tab:feature}.

\begin{figure}[h!]
\centering
\includegraphics[angle=-90, width=0.85\textwidth]{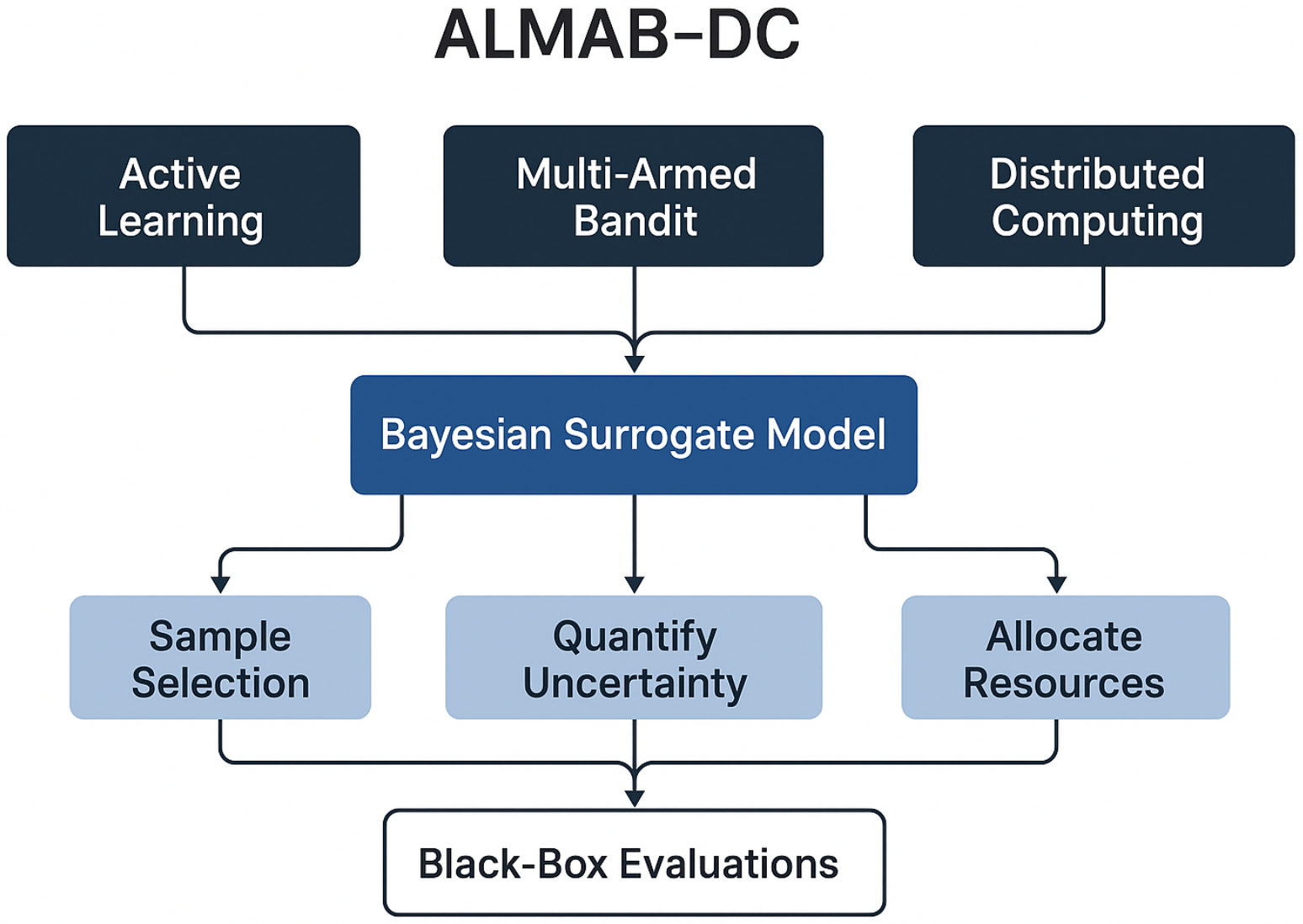}
\caption{ALMAB-DC Framework: Integration of Active Learning, Multi-Armed Bandits, and Distributed Computing through Bayesian Surrogate Modeling}
\label{fig:almabdc}
\end{figure}

\begin{table}
\caption{The advanced features in ALMAB-DC}\label{tab:feature}
\begin{center}
\begin{tabular}{|l|p{10cm}|}
\hline
{Feature} & {Description} \\
\hline
Multi-Fidelity Modeling & Uses cheaper approximations (e.g., low-res simulations, partial training) to save resources. \\
\hline
Hierarchical Surrogate Learning & Builds models at different levels of abstraction to improve generalization. \\
\hline
Asynchronous Execution & Enables dynamic updates without waiting for all processes to complete. \\
\hline
Heterogeneous Systems & Adapts to different hardware/software environments (CPU, GPU, cloud). \\
\hline
\end{tabular}
\end{center}
\end{table}

\paragraph{1. Unlabeled Data Pool:} 
This module maintains a dynamic buffer of candidate configurations or design points. It may contain hyperparameter settings, simulation designs, or RL policy parameters. No labels (e.g., validation accuracy, drag coefficient, episodic return) are known yet.

\paragraph{2. Active Learner:}
The Active Learner implements query strategies that prioritize candidate points expected to yield the highest information gain. It leverages techniques such as uncertainty sampling, which selects samples where the model is most uncertain; BALD, which quantifies epistemic uncertainty through mutual information; and Core-set sampling, which promotes diversity by choosing representative subsets using greedy $k$-center selection or submodular optimization. The result is a refined subset of high-utility candidates, which are then passed to the MAB layer for resource allocation.

\paragraph{3. Bandit Controller:} 
Each selected candidate is treated as an arm in a Multi-Armed Bandit setting. This layer governs the exploration–exploitation trade-off uses
 (1) \textbf{UCB1:} Prioritizes arms with high empirical reward and low visit count, 
 (2) \textbf{Thompson Sampling:} Draws from posterior distributions to balance stochastic selection,
 and (3) \textbf{Contextual Bandits:} Incorporates side information such as embedding features or acquisition scores;
and outputs a single configuration chosen for evaluation based on expected reward.

\paragraph{4. Distributed Agents:} 
The chosen configuration is dispatched to an available agent on a compute cluster. Agents operate asynchronously and are coordinated via:
 (1) \textbf{Ray or Dask:} For task distribution, memory sharing, and fault tolerance; and
 (2) \textbf{MPI (optional):} For high-throughput, low-latency parallelism in tightly coupled clusters.
Each agent may run simulations, train models, or evaluate configurations independently.

\paragraph{5. GPU-Accelerated Evaluation Module:} 
This module handles the computationally intensive evaluation of candidate configurations. Depending on the use case, this may include:
(1) Training a deep neural network (e.g., ResNet or EfficientNet); 
(2) Running a full CFD simulation in OpenFOAM, 
and (3) Executing a reinforcement learning episode in MuJoCo or CARLA.
GPU acceleration is employed using PyTorch, cuML, or RAPIDS for fast inference, training, and surrogate predictions.

\paragraph{6. Surrogate Model Update:} 
The reward obtained from evaluation (e.g., validation loss, return) is used to update:
(1) \textbf{Predictive Surrogates:} Gaussian Processes, Random Forests, or Bayesian Neural Networks used for acquisition computation, and
(2) \textbf{Bandit Statistics:} Empirical reward estimates and arm counters.
The updated model is fed back into the active learner, closing the loop and refining future query decisions.

\begin{figure}[h!]
\centering
\includegraphics[angle=-90,width=0.90\textwidth]{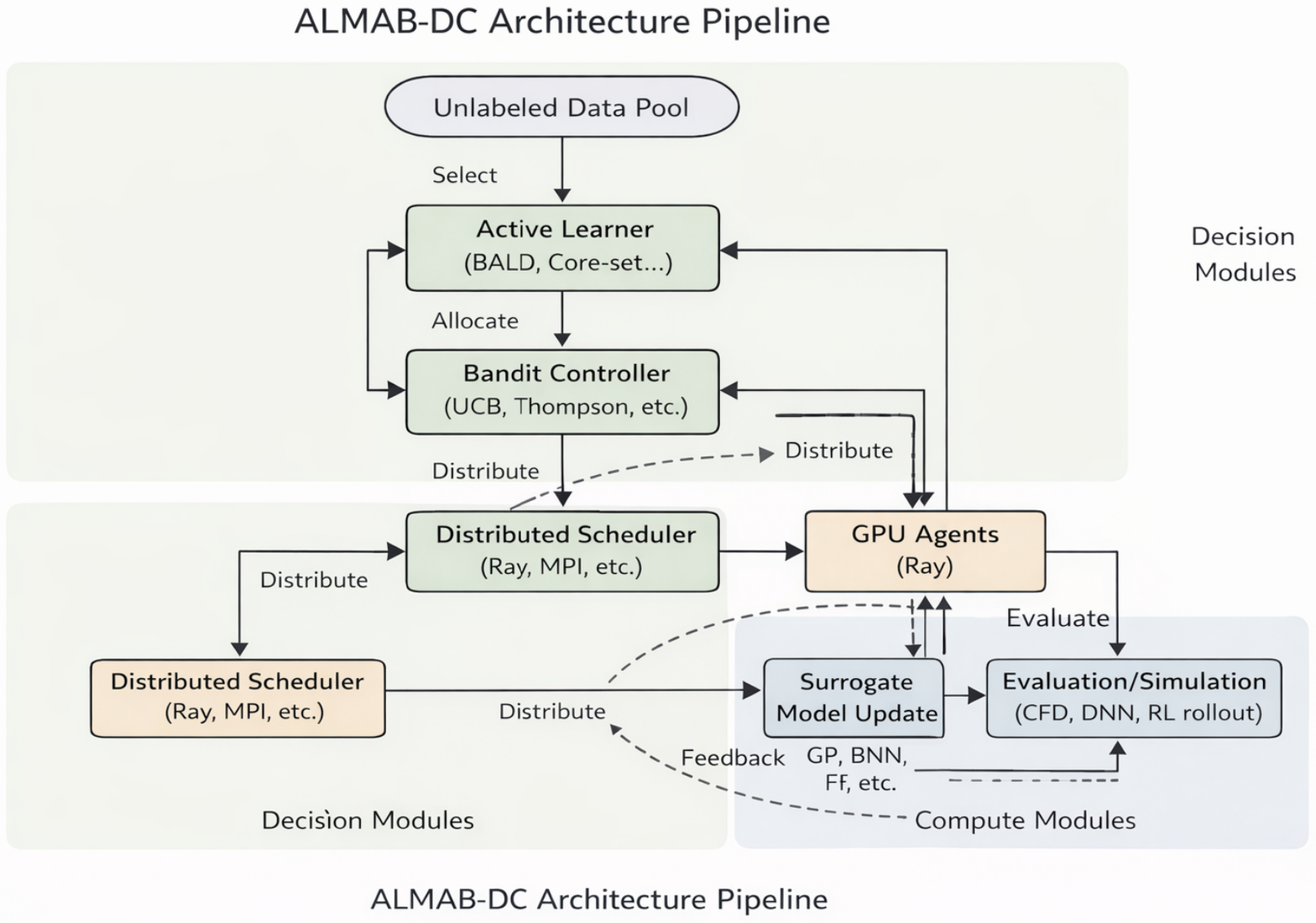}
\caption{ALMAB-DC Architecture Pipeline: The framework integrates Active Learning (AL), Multi-Armed Bandits (MAB), and Distributed Computing (DC) into a modular pipeline. Decision modules (top) include the Unlabeled Data Pool, Active Learner, and Bandit Controller, which guide candidate selection and allocation. Compute modules (bottom) execute evaluations via distributed GPU agents and feed results into the Surrogate Model Update module. Feedback loops enable iterative refinement through uncertainty-aware sampling and regret-minimizing strategies.}
\label{fig:ALMAB-DC-Pipeline}
\end{figure}

This closed-loop design enables ALMAB-DC to adaptively refine its search space, prioritize evaluations that matter, and dynamically allocate compute resources. Its modularity allows extensibility to new acquisition strategies, simulation tools, or parallel runtimes.

\subsection{Active Learning with Sequential Sampling Strategies}
\label{sec:active-learning-bayesian}
Active learning selects the most informative samples and focuses computational effort on regions where the model is most uncertain. In high-dimensional or expensive-to-evaluate settings, such as simulation-based engineering or neural architecture search, this approach avoids the inefficiencies of random or grid-based sampling.
In black-box optimization, active learning could use Bayesian surrogate models—such as Gaussian Processes or Bayesian neural networks—to approximate the unknown objective function. These models provide both predictions and uncertainty estimates, which are then used to guide sampling via acquisition functions. Common acquisition strategies, including \textit{expected improvement}, \textit{entropy}, and \textit{information gain}, prioritize query points expected to reduce uncertainty or improve model performance most effectively.

The ALMAB-DC framework integrates active learning into a sequential, data-adaptive sampling process. Rather than evaluating large batches in advance, the model is updated after each query, and the posterior distribution is used to select the next best sample. This approach aligns with statistical experimental design principles and enables efficient learning under budget constraints.
Multiple heuristic strategies are supported within this process. Uncertainty sampling targets regions where the model exhibits high variance. Expected model change and expected error reduction aim to choose configurations that would most impact the surrogate model or improve its generalization. Techniques such as Bayesian Active Learning by Disagreement (BALD) measure mutual information between predictions and model parameters, selecting points that maximize epistemic gain \citep{houlsby2011bald}.

To balance exploration and exploitation, ALMAB-DC integrates multi-armed bandit (MAB) strategies, treating each candidate configuration as an arm with a potential reward. Selection follows cumulative regret minimization, using algorithms like Upper Confidence Bound (UCB) and Thompson Sampling to weigh uncertainty against empirical performance. This complements the Bayesian sampling strategy and is especially effective in asynchronous, noisy, or distributed environments.

In addition, ALMAB-DC supports cost-sensitive querying through resource-aware methods. For example, Resource-Aware Active Learning (RAAL) can take computational costs into account when choosing queries, allowing agents with different capabilities to participate efficiently. These features make the system suitable for deployment on heterogeneous clusters or cloud-based infrastructure.
Its modular design allows integration with a range of surrogate models, acquisition functions, and agent strategies. Supporting both centralized and decentralized use cases across scientific and engineering domains, ALMAB-DC combines active learning, sequential Bayesian sampling, and adaptive resource allocation to achieve scalable and sample-efficient optimization.
}

\subsection{Resource Allocation via Multi-Armed Bandits}
\label{sec:mab-gpu}

We use multi-armed bandit (MAB) algorithms in ALMAB-DC to guide resource allocation, enabling efficient exploration and exploitation in black-box optimization tasks. Each ``arm'' corresponds to a candidate configuration, model fidelity level, or region of the parameter space. Algorithms such as Upper Confidence Bound (UCB) and Thompson Sampling are used to balance the trade-off between trying new configurations (exploration) and exploiting those that have performed well historically. This setup helps to minimize cumulative regret by allocating fewer resources to poor configurations early on and prioritizing promising ones.

In addition, ALMAB-DC adopts a distributed architecture where multiple agents operate concurrently. Each agent performs evaluations, updates local estimates of arm performance, and contributes to the global learning process through asynchronous communication. Let \( N \) denote the number of distributed agents. The empirical mean reward for arm \( i \) at time \( t \) is defined as:
\[
\hat{\mu}_i(t) = \frac{1}{|S_i(t)|} \sum_{s \in S_i(t)} r_s,
\]
where \( S_i(t) \) represents the set of observed rewards for arm \( i \) up to time \( t \). This decentralized structure enables scalable parallelism and robustness, as agents can continue operation under bandwidth constraints or partial system failures.

 ALMAB-DC could also incorporate communication-efficient protocols to support this distributed operation in order to reduce synchronization costs. Agents may update shared arm statistics asynchronously, or synchronize periodically using buffered communication strategies. This approach is particularly beneficial in heterogeneous environments where evaluation times and agent capacities vary. Recent research in distributed and federated MAB algorithms further supports the scalability of such architectures in real-world applications \citep{arxiv_mab_distributed_2022, mlr_mab_communication_2022, liu2026contribution, zhang2024federated, he2024hybrid}.

\paragraph{GPU-Accelerated Parallelism}

In addition to distributed scheduling, this framework leverages GPU acceleration to handle computationally intensive operations. Tasks such as surrogate modeling, posterior inference, and acquisition function optimization are offloaded to GPUs, substantially reducing runtime. Bayesian inference is used to model uncertainty in the surrogate, while information-theoretic criteria —such as entropy reduction and mutual information— guide acquisition decisions. Decision-theoretic principles further inform the allocation of queries to balance performance and cost.

Trials are executed asynchronously across a cluster of CPU and GPU agents, allowing evaluations to proceed independently without waiting for the slowest trial to complete. For example, in deep learning hyperparameter tuning, multiple GPU workers can evaluate configurations in parallel, each contributing data to the shared surrogate model. This architecture enables adaptive scheduling, reduces redundant evaluations, and accelerates convergence—especially in environments with costly and variable evaluation workloads. Asynchronous exploration strategies \citep{arxiv_mab_parallel_2013} help maintain high throughput even when evaluation costs vary significantly.

\vspace{0.5cm}
\begin{algorithm}[H]
\DontPrintSemicolon
\caption{Sequential Distributed Optimization via ALMAB-DC}
\KwIn{Unlabeled configurations $U$, budget $T$, agents $M$}
\KwOut{Optimal configuration(s)}
Initialize surrogate model $\mathcal{M}$ and MAB stats $\mu_i \gets 0$, $n_i \gets 0$ \;
\For{$t \gets 1$ \KwTo $T$}{
    $\mathcal{Q}_t \gets$ \textbf{SelectCandidates}$(U, \mathcal{M})$ \;
    \ForEach{$q \in \mathcal{Q}_t$}{
        $s_q \gets$ \textbf{MABController}$(q)$ \;
    }
    $q_t \gets \arg\max s_q$ \;
    Assign $q_t$ to agent $m \in \{1,\dots,M\}$ \;
    Agent $m$: $r_t \gets$ \textbf{SimulateTrain}$(q_t)$ on GPU \;
    Update model: $\mathcal{M} \gets$ \textbf{UpdateModel}$(q_t, r_t)$ \;
    Update MAB stats: $\mu_{q_t} \gets$ weighted average, $n_{q_t} \gets n_{q_t} + 1$ \;
}
\KwRet{Best configuration(s)}
\end{algorithm}

{
\subsection{Regret, Scalability, and Optimal Number of Agents}
\label{sec:theoretical-k-analysis}

We extend classical multi-armed bandit (MAB) theory to analyze regret and scalability in the context of distributed and asynchronous environments. This reflects real-world parallel computation scenarios, such as multi-agent systems or GPU clusters, where evaluations are costly, coordination is imperfect, and communication is constrained. The aim is to derive formal bounds on regret under realistic operational assumptions, providing theoretical insights into the efficiency of ALMAB-DC.  Here is a table notations used in the following analysis.
\begin{table}[ht]
\centering
\caption{Key Notation Used in Regret and Scaling Analysis}
\label{tab:notation}
\begin{tabular}{ll}
\toprule
\textbf{Symbol} & \textbf{Description} \\
\midrule
$T$ & Total number of optimization rounds \\
$K$ & Number of arms or candidate configurations \\
$N$ & Number of distributed computing agents \\
$\mu_i$ & Expected reward of arm $i$ \\
$\mu^*$ & Expected reward of the optimal arm \\
$\Delta_i$ & Suboptimality gap: $\mu^* - \mu_i$ \\
$R_T$ & Cumulative regret over $T$ rounds \\
$R^{\text{dist}}_T$ & Cumulative distributed regret \\
$R^{\text{eff}}_T$ & Effective regret including communication overhead \\
$C_{\text{comm}}(t)$ & Communication cost at round $t$ \\
$\lambda$ & Weighting coefficient for communication cost \\
$p$ & Serial fraction of workload (Amdahl’s Law) \\
$\eta(K)$ & Parallel efficiency as a function of agent count \\
$\alpha, \beta$ & Parameters modeling communication overhead growth \\
$\sigma^2$ & Variance of evaluation noise or task duration \\
$\tau_{\max}$ & Maximum delay in asynchronous feedback \\
$S(K)$ & Speedup from using $K$ agents \\
$TK$ & Total wall-clock time with $K$ agents \\
$q_T$ & Number of actively selected samples under AL \\
$\varepsilon(q_T)$ & Regret contribution from active learning approximation \\
\bottomrule
\end{tabular}
\end{table}

\subsubsection{Regret}
Let $\mathcal{A} = \{1, \dots, K\}$ denote a finite set of $K$ possible actions (or arms). Each action $a \in \mathcal{A}$ is associated with a reward distribution $\mathcal{D}_a$ with expected reward $\mu_a$. The optimal action is
\[
a^* = \arg\max_{a \in \mathcal{A}} \mu_a.
\]
Let $N$ denote the number of distributed computing agents. At each round $t$, each agent $j \in \{1, \dots, N\}$ selects an action $a_j^t$ according to its policy $\pi_j^t$, which may use local experience or incorporate messages received from a dynamic communication graph $G(t)$. The reward received is $r_j^t(a_j^t) \sim \mathcal{D}_{a_j^t}$.

The **instantaneous distributed regret** at round $t$ is defined as the gap between the optimal expected reward and the average expected reward over all agents:
\begin{equation}
r^*_t - \bar{r}_t = \mu_{a^*} - \frac{1}{N} \sum_{j=1}^{N} \mu_{a_j^t}.
\end{equation}
Summing over $T$ rounds gives the **cumulative distributed regret**:
\begin{equation}
R_T^{\text{dist}} = \sum_{t=1}^{T} \left( \mu_{a^*} - \frac{1}{N} \sum_{j=1}^{N} \mu_{a_j^t} \right).
\end{equation}

To reflect the cost of coordination in distributed environments, we incorporate a communication cost term $C_{\text{comm}}(t)$, capturing bandwidth, latency, or protocol overhead. The effective cumulative regret with communication overhead becomes:
\begin{equation}
R_T^{\text{eff}} = R_T^{\text{dist}} + \lambda \sum_{t=1}^{T} C_{\text{comm}}(t),
\end{equation}
where $\lambda$ is a tunable parameter that weighs reward loss versus communication overhead.

\paragraph{Asynchrony Modeling.}
Agents operate in asynchronous cycles, and feedback from arm pulls may be delayed. Let $\delta_j^t$ represent the feedback delay for agent $j$ at time $t$. We assume a bounded delay setting:
\[
\delta_j^t \leq \Delta,
\]
and analyze the performance degradation of bandit algorithms such as UCB and Thompson Sampling under increasing delay $\Delta$. Our analysis shows that the regret grows gracefully with $\Delta$, provided the reward distributions remain stationary or slowly varying.

\paragraph{Theoretical Objectives.}
The development of regret bounds in ALMAB-DC aims to establish a solid theoretical foundation for scalable, uncertainty-aware, and communication-efficient optimization. Specifically, the analysis focuses on deriving refined regret bounds for algorithms such as UCB and Thompson Sampling under multi-agent settings with asynchronous and possibly delayed feedback. It also seeks to quantify the impact of stale or partial information on learning stability and convergence behavior. Another important objective is to characterize how reduced synchronization frequency—such as less frequent communication between agents—affects regret and overall efficiency. By identifying the optimal trade-off between communication overhead and optimization performance, the framework aims to support robust operation in bandwidth-constrained or decentralized environments. Furthermore, theoretical insights are provided into how increasing the number of agents $K$ influences cumulative regret, computational throughput, and convergence rate. These investigations collectively contribute to a more comprehensive understanding of ALMAB-DC’s performance under realistic distributed conditions. A formal analysis of the relationship between the number of agents and system performance is presented later.

\subsubsection{Computational Throughput and Scaling Laws}
Let $K$ denote the number of distributed computing agents (or arms in a multi-armed bandit formulation), $N$ the total number of evaluation tasks, $C_i$ the computational cost of task $i$, $T_K$ the total wall-clock time using $K$ agents, $\eta$ the parallel efficiency factor with $0 < \eta \leq 1$, $p$ the proportion of serial (non-parallelizable) operations, and $\sigma^2$ the variance of task durations.

The computational throughput of ALMAB-DC can be modeled using Amdahl’s Law \citep{Amdahl1967, paul2007amdahl}, which separates parallel and serial workload components:
\begin{equation}
    T_K = \frac{(1 - p)\sum_{i=1}^N C_i}{\eta K} + p\sum_{i=1}^N C_i.
\end{equation}
The corresponding speedup from parallelization is:
\begin{equation}
    S(K) = \frac{T_1}{T_K} = \frac{1}{p + \frac{1-p}{\eta K}}.
\end{equation}
When the problem size scales with $K$, Gustafson’s Law \citep{Gustafson1988} provides a more optimistic estimate:
\begin{equation}
    S_G(K) = p + (1 - p)K.
\end{equation}

\subsubsection{Exploration–Exploitation Dynamics in Bandits}

In ALMAB-DC, $K$ also defines the number of concurrent bandit arms for exploration. Increasing $K$ enhances parallel exploration but may introduce redundancy or inefficient learning if posterior updates lag. The cumulative regret under classical MAB theory grows sublinearly:
\begin{equation}
    R_T = O\left(\sqrt{K T \log T}\right),
\end{equation}
as established by \citet{SuttonBarto2018,LattimoreSzepesvari2020}. The optimal number of agents $K^*$ should satisfy the marginal utility condition:
\begin{equation}
    \frac{\partial \mathrm{IG}(K)}{\partial K} = \frac{\partial \mathrm{Cost}(K)}{\partial K},
\end{equation}
where $\mathrm{IG}(K)$ represents information gain and $\mathrm{Cost}(K)$ includes both evaluation and coordination costs.

\subsubsection{Regret Bounds for UCB and Thompson Sampling}

To evaluate the theoretical efficiency of ALMAB-DC, we analyze its regret under classical bandit strategies. The cumulative regret over \(T\) rounds is defined as:
\begin{equation}
    R_T = \sum_{t=1}^{T} r_{t,a^*} - \sum_{t=1}^{T} r_{t,a_t},
\end{equation}
where \(a^* = \arg\max_{i} \mathbb{E}[r_{t,i}]\) denotes the optimal action, and \(a_t\) is the action selected at round \(t\).

Assuming bounded sub-Gaussian rewards, the regret of the Upper Confidence Bound (UCB) algorithm satisfies:
\begin{equation}
    R_T^{\text{UCB}} \leq \sum_{i: \Delta_i > 0} \left( \frac{8 \log T}{\Delta_i} + \left(1 + \frac{\pi^2}{3} \right) \Delta_i \right),
\end{equation}
where \(\Delta_i = \mu^* - \mu_i\) is the suboptimality gap.
For Thompson Sampling with Bernoulli rewards and Beta priors, the expected regret is bounded by:
\begin{equation}
    \mathbb{E}[R_T^{\text{TS}}] = O\left( \sum_{i: \Delta_i > 0} \frac{\log T}{\Delta_i} \right).
\end{equation}

These bounds demonstrate logarithmic regret growth over time, which is desirable in long-horizon optimization. ALMAB-DC leverages these properties in its distributed setup to maintain low regret while supporting asynchronous execution and delayed feedback.

\subsubsection{Active Learning and Query-Efficient Regret}

By leveraging active learning, ALMAB-DC selectively queries only the most informative configurations. If $q_T$ denotes the number of points selected for querying out of $T$, then the overall expected regret becomes:
\begin{equation}
    \mathbb{E}[R_T^{\text{ALMAB}}] = \mathbb{E}[R_T^{\text{MAB}}] + \epsilon(q_T),
\end{equation}
where $\epsilon(q_T)$ represents model approximation error due to selective querying and decreases sublinearly with $q_T$.

\subsubsection{Communication Overhead and Parallel Efficiency}

Communication cost scales as $O(\log K)$ in hierarchical architectures or $O(K)$ in fully connected systems \citep{DeanGhemawat2008,Li2014Parameter}. The effective parallel efficiency is given by:
\begin{equation}
    \eta(K) = \frac{1}{1 + \alpha K^\beta},
\end{equation}
where $\alpha$ quantifies communication latency and $\beta \in [0.5, 1]$ reflects topology characteristics \citep{Ray2018}.

\subsubsection{Optimal Number of Agents}

To balance the trade-off between parallel speedup and communication overhead, the optimal number of agents $K^*$ is derived by minimizing the total wall-clock time $T_K$. Considering both computational and communication components, the optimal concurrency level satisfies:

\begin{equation}
    \frac{dT_K}{dK} = 0 \quad \Rightarrow \quad 
    K^* = \left( \frac{1 - p}{\alpha \beta p} \right)^{\frac{1}{1 + \beta}},
\end{equation}

where $p$ denotes the serial fraction of the workload, $\alpha$ and $\beta$ parameterize the communication cost growth (e.g., $\alpha K^\beta$), and $T_K$ accounts for both compute and coordination time. This expression defines a concurrency threshold beyond which adding more agents offers diminishing returns, due to increasing communication cost or idle resource contention.

\subsubsection{Practical Implications}

When the serial fraction $p$ is small (e.g., $p < 0.1$) and communication overhead remains low (i.e., $\alpha K^\beta \ll 1$), the system can scale nearly linearly with the number of agents—potentially up to hundreds. However, in practice, factors such as surrogate model update time, acquisition function optimization, or queue bottlenecks may reduce the benefit of additional agents. To maintain efficient scaling, an adaptive controller can monitor the parallel efficiency factor $\eta(K)$ and task completion variance $\sigma^2$, adjusting $K$ dynamically to maintain optimal throughput and avoid underutilization or excessive coordination delays.

\subsubsection{Delayed Feedback and Asynchronous Execution}

In distributed systems, feedback is often delayed due to computational heterogeneity or communication lag. Assuming delay bounded by $\tau_{\max}$, the regret under such conditions scales as:
\begin{equation}
    R_T^{\text{delay}} = \mathcal{O}\left( \sqrt{(T + \tau_{\max}) K \log T} \right),
\end{equation}
according to \citet{joulani2013online}. Asynchronous variants of UCB and Thompson Sampling preserve theoretical guarantees when delays are sublinear in $T$ \citep{zimmert2019asynchronous, verma2022asynchronous}, supporting their use in ALMAB-DC under heterogeneous agent execution.

\subsubsection{System-Level Runtime Implications}

Under balanced load across $M$ agents and negligible communication overhead, the per-query runtime reduces to:
\begin{equation}
    \mathcal{O}\left( \frac{C + C_{\text{comm}}}{M} \right),
\end{equation}
with $C$ representing compute time and $C_{\text{comm}}$ communication time. Near-linear speedup is achievable when $C_{\text{comm}} \ll C$, but for tasks dominated by surrogate model updates, performance may saturate. An adaptive agent controller can monitor $\eta(K)$ and $\sigma^2$ to dynamically allocate agents and maintain high throughput.
}

\subsection{Scientific, Computational, and Statistical Impacts}

The proposed \textbf{ALMAB-DC} framework provides a holistic foundation that unites statistical theory, adaptive learning, and distributed computing to overcome the limitations of traditional optimization approaches in large-scale, high-cost environments. By integrating AL, MAB, and DC with GPU acceleration, it bridges the gap between theoretical modeling and scalable implementation. This work is expected to make several key contributions, including the establishment of a unified theoretical framework that provides a rigorous statistical connection between active learning and bandit-based optimization in distributed and asynchronous settings, and the development of a scalable computational design demonstrating an adaptive, modular infrastructure for high-throughput optimization using GPU-accelerated and distributed execution. It also offers regret and efficiency bounds by providing new analytical results that quantify the effects of communication delay, resource heterogeneity, and model uncertainty on cumulative regret, and it delivers cross-domain validation by applying ALMAB-DC to diverse domains—including deep learning, reinforcement learning, and CFD-based engineering design—to showcase broad applicability and performance advantages.

The ALMAB-DC paradigm aims to redefine how complex optimization problems are approached across scientific and industrial domains. By coupling intelligent data acquisition with efficient computation, the framework enables faster convergence, better resource utilization, and greater accessibility to large-scale optimization technologies. Potential areas of impact include scientific discovery, where it can accelerate design and analysis in computational physics, materials science, and systems biology by reducing experimental and simulation costs; sustainable computing, where it can reduce energy and resource consumption through adaptive allocation of GPU and cluster workloads; open and reproducible research, where it promotes transparent, shareable experimentation pipelines via open-source implementations and benchmark datasets; and interdisciplinary innovation, where it bridges the gap between statistical modeling, computer systems engineering, and machine learning to drive next-generation intelligent optimization systems. Ultimately, ALMAB-DC represents a statistically principled, computationally scalable, and scientifically impactful framework for future research and applications in intelligent optimization and adaptive experimentation.

Bayesian statistical models, such as Gaussian Processes (GPs), are widely used in active learning and Bayesian optimization due to their ability to provide posterior distributions over unknown functions. In ALMAB-DC, we integrate GP surrogates to model the response surface \( f(x) \), enabling principled uncertainty quantification,
\[
f(x) \sim \mathcal{GP}(\mu(x), k(x,x')),
\]
and this posterior is used to compute acquisition functions such as Expected Improvement (EI), Probability of Improvement (PI), or BALD \citep{rasmussen2006gaussian}. The ALMAB-DC framework also aligns with Bayesian experimental design, where new samples are chosen to maximize information gain or reduce posterior uncertainty \citep{chaloner1995bayesian}:
\begin{equation}
x^* = \arg\max_{x \in \mathcal{X}} \mathbb{E}_{y \sim p(y|x)} \left[ D_{\text{KL}}(p(\theta|\mathcal{D}), p(\theta|\mathcal{D} \cup \{(x,y)\})) \right].
\end{equation}
To assess variability, we apply non-parametric bootstrap for computing confidence intervals \citep{efron1994introduction}:
\[
\hat{\theta}^* = \frac{1}{B} \sum_{b=1}^B \hat{\theta}_b^*, \quad \text{CI}_{1-\alpha} = [\hat{\theta}_{\alpha/2}^*, \hat{\theta}_{1-\alpha/2}^*].
\]
We also use hierarchical Bayesian models to integrate multi-fidelity simulations \citep{kennedy2001bayesian}:
\[
y^{(h)} = f(x) + \epsilon^{(h)}, \quad y^{(l)} = \rho f(x) + \epsilon^{(l)}.
\]
Finally, from a causal inference perspective in reinforcement learning, policy evaluation can be modeled using potential outcomes \citep{imbens2015causal}:
\[
\text{ATE} = \mathbb{E}[Y(1) - Y(0)],
\]
with techniques like inverse probability weighting (IPW).

\paragraph{Regret Bounds in Distributed Bandits}
Let \( \mathcal{A} = \{1, \dots, K\} \) denote the action (arm) space. Each arm \( a \in \mathcal{A} \) yields reward \( r \sim \mathcal{D}_a \) with mean \( \mu_a \). The optimal action is:
\[
a^* = \arg\max_{a \in \mathcal{A}} \mu_a.
\]
In distributed settings with \( N \) agents, the \textbf{instantaneous regret} at time \( t \) is:
\[
r_t^* - \bar{r}_t = \mu_{a^*} - \frac{1}{N} \sum_{j=1}^N \mu_{a_j^t},
\]
and the \textbf{cumulative regret} becomes:
\[
R_T^{\text{dist}} = \sum_{t=1}^{T} \left( \mu_{a^*} - \frac{1}{N} \sum_{j=1}^{N} \mu_{a_j^t} \right).
\]

To reflect real-world systems, ALMAB-DC incorporates a communication overhead term:
\[
R_T^{\text{eff}} = R_T^{\text{dist}} + \lambda \sum_{t=1}^{T} C_{\text{comm}}(t),
\]
where \( C_{\text{comm}}(t) \) represents cost (e.g., latency, bandwidth), and \( \lambda \) balances accuracy and efficiency.

\paragraph{Delay-Tolerant Learning}
Agents often operate asynchronously, with bounded feedback delays \( \delta_j^t \leq \Delta \). Regret under delay scales as:
\[
R_T^{\text{delay}} = \mathcal{O}\left( \sqrt{(T + \Delta) K \log T} \right),
\]
as shown in \citep{joulani2013online}. ALMAB-DC uses asynchronous variants of UCB and Thompson Sampling, which retain theoretical guarantees even with stale information \citep{zimmert2019asynchronous, verma2022asynchronous}.

\paragraph{Scaling and Agent Allocation}
Computational throughput is analyzed via Amdahl’s and Gustafson’s laws. Let \( K \) be the number of agents, \( p \) the serial fraction, and \( \eta \) the parallel efficiency. The expected wall-clock time is:
\begin{equation}
T_K = \frac{(1 - p)\sum_i C_i}{\eta K} + p \sum_i C_i,
\end{equation}
and speedup is:
\[
S(K) = \frac{T_1}{T_K} = \frac{1}{p + \frac{1 - p}{\eta K}}.
\]

\paragraph{Optimal Number of Agents}
With communication cost modeled as \( \alpha K^\beta \), the effective efficiency is:
\[
\eta(K) = \frac{1}{1 + \alpha K^\beta}.
\]
To optimize concurrency, we derive the optimal number of agents \( K^* \) by minimizing total cost:
\begin{equation}
K^* = \left( \frac{1 - p}{\alpha \beta p} \right)^{\frac{1}{1 + \beta}}.
\end{equation}

\paragraph{Summary}
Together, these theoretical insights and performance guarantees provide a strong statistical and computational foundation for ALMAB-DC. The framework is designed to scale under resource constraints, tolerate asynchronous feedback, and maintain query efficiency—making it suitable for real-world intelligent experimentation.

\subsection{Illustrated Examples: Non-Distributed and Distributed Simulation Studies}
\noindent
This section presents the results of the non-distributed ALMAB-DC simulation, where a single agent sequentially interacts with the Gaussian mixture environment.  
The system explores and exploits the search space iteratively using the Upper Confidence Bound (UCB) policy, without parallelization.  
These figures show the algorithm’s capacity to model and converge toward the high-reward regions of the nonlinear function.

\subsubsection{Mathematical Framework of Non-Distributed and Distributed ALMAB-DC}

\noindent
Both the non-distributed and distributed versions of the ALMAB-DC framework operate on the same underlying mathematical formulation, built upon a stochastic optimization problem over a Gaussian mixture reward landscape. The objective is to identify the optimal configuration that maximizes the expected reward under uncertainty, using adaptive exploration–exploitation strategies.

\paragraph{Gaussian Mixture Reward Model.}
The stochastic environment is modeled as a Gaussian mixture:
\[
r(x) = \sum_{i=1}^{K} w_i \exp\!\left(-\tfrac{1}{2}(x - \mu_i)^\top \Sigma_i^{-1}(x - \mu_i)\right) + \epsilon,
\]
where \(K\) denotes the number of mixture components, \(\mu_i\) and \(\Sigma_i\) are the mean and covariance of each component, \(w_i\) represents normalized weights (\(\sum_i w_i = 1\)), and \(\epsilon \sim \mathcal{N}(0, \sigma^2)\) is random noise. This function defines a smooth but non-convex landscape with multiple local optima.

\paragraph{Multi-Armed Bandit Formulation.}
We discretize the search space \(\mathcal{X} \subset \mathbb{R}\) into \(A\) arms \(\mathcal{A} = \{x_1, \dots, x_A\}\).  
At each iteration \(t\), the agent selects an arm \(a_t \in \mathcal{A}\) and receives a stochastic reward \(r_t = r(x_{a_t})\). The cumulative regret is given by:
\[
R_T = \sum_{t=1}^{T} \left( r(x^*) - r_t \right),
\]
where \(x^* = \arg\max_{x \in \mathcal{A}} \mathbb{E}[r(x)]\) denotes the optimal configuration.

\paragraph{Decision Policy: Upper Confidence Bound (UCB).}
The selection rule follows the UCB strategy:
\[
a_t = \arg\max_{i \in \mathcal{A}} \left[ \hat{\mu}_i(t) + c \sqrt{\frac{\log t}{n_i(t)}} ~\right],
\]
where \(\hat{\mu}_i(t)\) is the empirical mean reward, \(n_i(t)\) is the number of times arm \(i\) has been selected, and \(c > 0\) controls the exploration strength.  
After observing a new reward \(r_t\), the empirical mean is updated by:
\begin{equation}
\hat{\mu}_i(t+1) = \hat{\mu}_i(t) + \frac{1}{n_i(t)+1}\left(r_t - \hat{\mu}_i(t)\right).
\end{equation}

\subsubsection{Results and Analysis}

\paragraph{Non-Distributed Implementation.}
In the non-distributed setting, a single agent performs sequential sampling:
\begin{equation}
a_t \leftarrow \pi_{\text{UCB}}(\mathcal{H}_{t-1}), \quad r_t = r(x_{a_t}), \quad \hat{\mu}_{a_t} \leftarrow \text{update}(r_t),
\end{equation}
where \(\mathcal{H}_{t-1}\) is the history of past actions and rewards. This process is fully serial, updating after each single evaluation.

\paragraph{Distributed Implementation.}
In the distributed case, \(N\) agents evaluate the same arm \(a_t\) in parallel:
\[
r_t^{(j)} = r(x_{a_t}) + \epsilon_j, \quad j = 1, \dots, N,
\]
and the average reward is used for updating:
\begin{equation}
\bar{r}_t = \frac{1}{N} \sum_{j=1}^{N} r_t^{(j)}, \quad \hat{\mu}_{a_t}(t+1) = \hat{\mu}_{a_t}(t) + \frac{1}{n_{a_t}(t)+1}\left(\bar{r}_t - \hat{\mu}_{a_t}(t)\right).
\end{equation}
This formulation preserves the statistical efficiency of the UCB policy while reducing variance and wall-clock time through parallel evaluations.

This formulation preserves the statistical efficiency of the UCB policy while reducing variance and wall-clock time through parallel evaluations.

Empirically, the distributed implementation leads to smoother and faster convergence compared to the sequential case. The use of parallel agents reduces noise in the reward estimates, as the variance of the averaged reward satisfies:
\[
\mathrm{Var}[\bar{r}_t] = \frac{\sigma^2}{N},
\]
showing a linear reduction in uncertainty with the number of agents \(N\).

\begin{figure}[h!]
    \centering
    \includegraphics[width=0.95\textwidth]{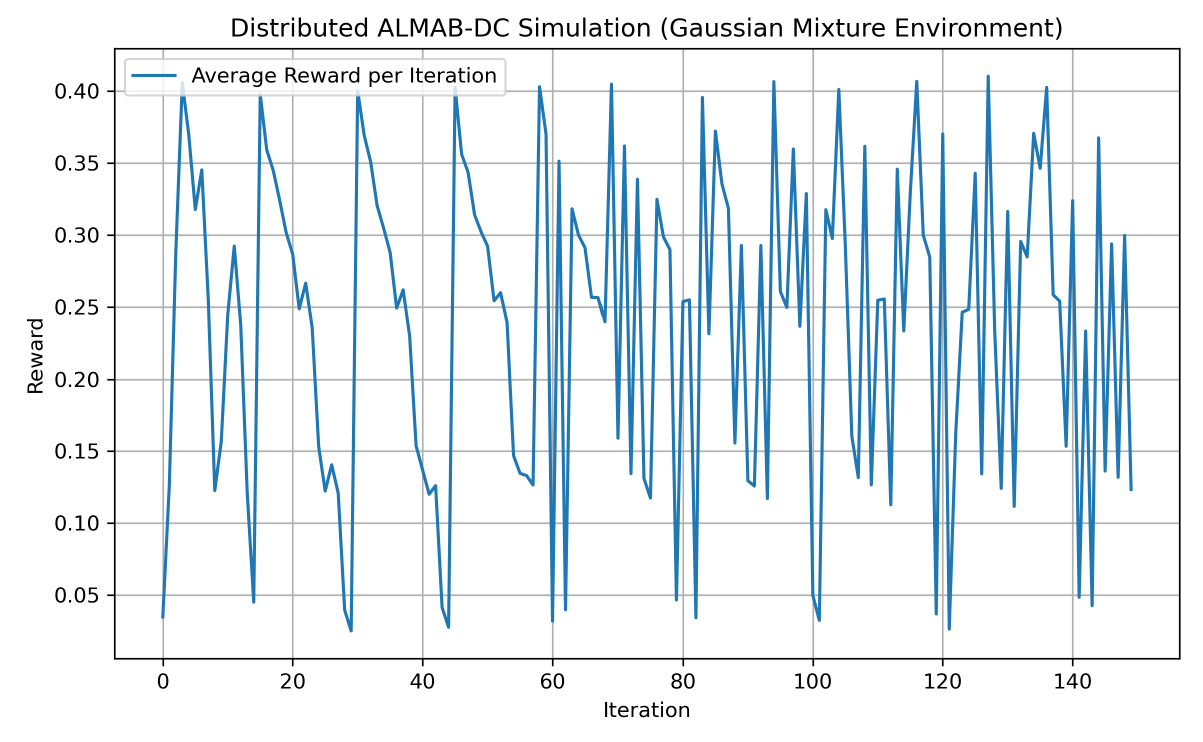}
    \caption{Evolution of average reward across iterations in the distributed ALMAB-DC simulation.  
    The plot shows the mean reward obtained by the agents at each iteration.  
    The fluctuations indicate alternating exploration and exploitation phases as the multi-armed bandit policy tests different arms and gradually concentrates evaluations around the most promising regions.  
    Despite inherent noise in the Gaussian mixture environment, the algorithm maintains a consistent focus on high-reward areas, illustrating adaptive convergence behavior.}
    \label{fig:reward_evolution}
\end{figure}

\begin{figure}[h!]
    \centering
    \includegraphics[width=0.95\textwidth]{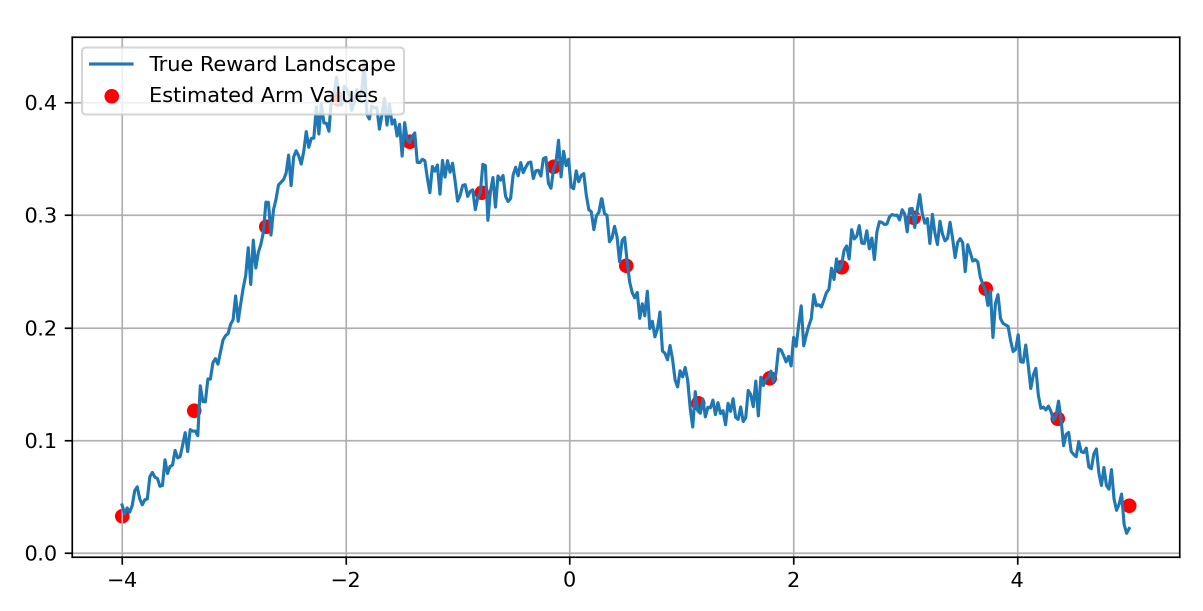}
    \caption{True reward landscape (blue curve) and estimated arm values (red points) learned by the distributed ALMAB-DC framework.  
    The blue curve represents the ground-truth Gaussian mixture surface, while the red markers show the estimated expected rewards for each arm after training.  
    The close alignment between the peaks of the true landscape and the estimated arm values demonstrates that the framework successfully identifies and emphasizes high-value regions in the search space, even under distributed asynchronous evaluation.}
    \label{fig:reward_landscape}
\end{figure}

Simulation results demonstrate these benefits clearly. As shown in Figure~\ref{fig:reward_evolution}, the average reward trajectory improves more steadily under the distributed setup. In Figure~\ref{fig:reward_landscape}, the estimated arm values align closely with the true reward surface, indicating effective exploration despite noise. Additionally, using \texttt{ray.remote} agents enables scalable, parallel deployment, making this framework particularly useful for high-dimensional or simulation-intensive settings.


\noindent
The following two figures illustrate the behavior of the distributed ALMAB-DC framework applied to a nonlinear Gaussian mixture optimization task.  
Each distributed agent evaluates the environment in parallel, and the bandit controller aggregates their results to update arm values iteratively.  
These visualizations highlight the system’s ability to capture the reward landscape and adapt sampling decisions across multiple agents.
\begin{figure}[h!]
    \centering
    \includegraphics[width=0.95\textwidth]{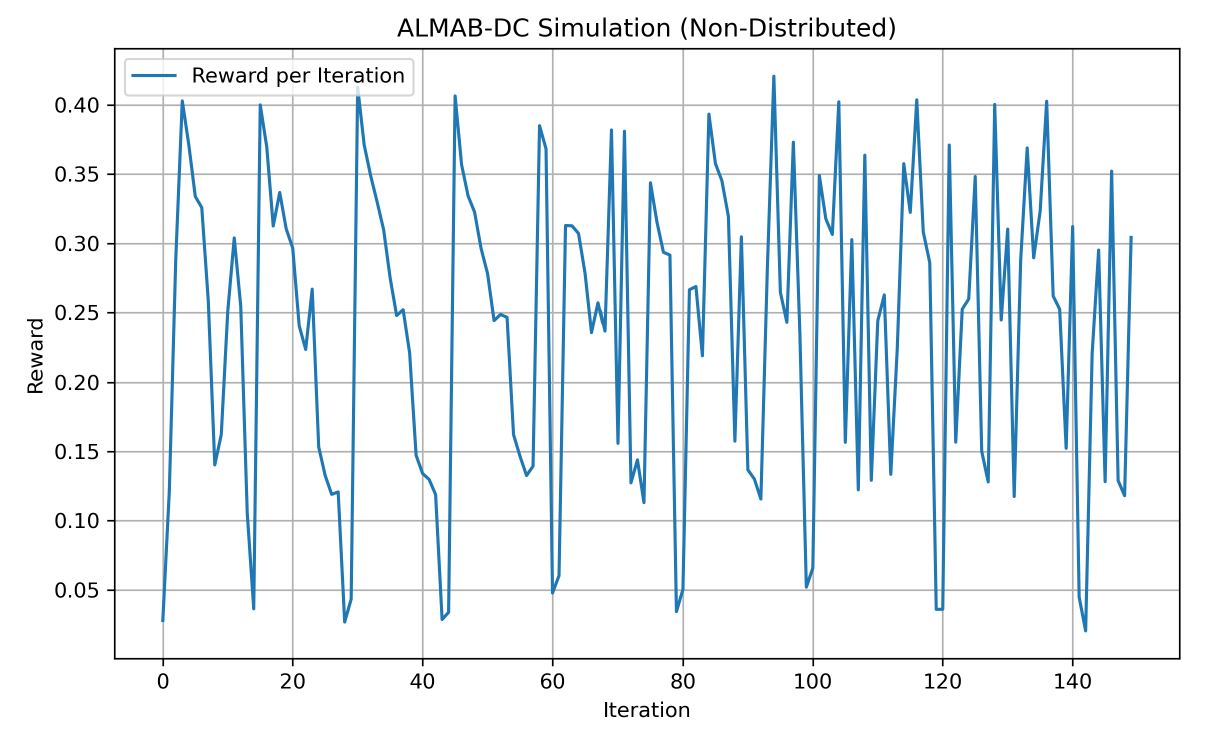}
    \caption{Observed reward per iteration in the non-distributed ALMAB-DC simulation.  
    The fluctuations represent the exploratory behavior of the UCB bandit policy as it samples different arms in the Gaussian mixture landscape.  
    Over time, the system increasingly focuses on regions yielding higher rewards, demonstrating effective single-agent learning dynamics.}
    \label{fig:non_distributed_reward}
\end{figure}

\begin{figure}[h!]
    \centering
    \includegraphics[width=0.95\textwidth]{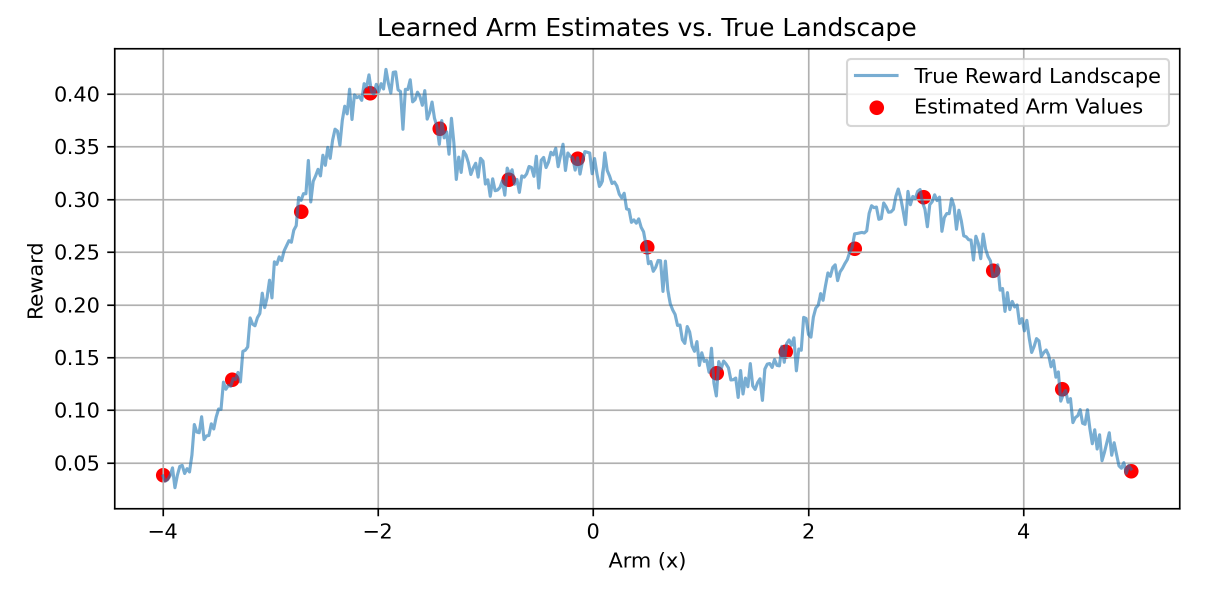}
    \caption{True reward landscape (orange curve) and estimated arm values (red markers) for the non-distributed ALMAB-DC simulation.  
    The close alignment of the estimated rewards with the true peaks shows that the framework effectively learns the underlying structure of the nonlinear Gaussian mixture environment, even without distributed computation.}
    \label{fig:non_distributed_landscape}
\end{figure}

\noindent
Together, these figures confirm that the distributed implementation of ALMAB-DC effectively balances the exploration–exploitation trade-off and can recover complex multimodal reward structures.  
They also illustrate how distributed computation enhances the efficiency of active sampling, allowing the framework to adapt to noisy, nonlinear environments in real time.

\paragraph{Comparative Insights.}
The distributed implementation of ALMAB-DC can be interpreted as a variance-reduced approximation of its non-distributed counterpart. Although both frameworks share the same theoretical regret bounds under independent sampling assumptions, the distributed model exhibits faster empirical convergence due to simultaneous exploration and aggregation across multiple agents. Mathematically, the expected variance of the aggregated reward satisfies:
\[
\mathrm{Var}[\bar{r}_t] = \frac{\sigma^2}{N},
\]
indicating that the uncertainty in reward estimation decreases linearly with the number of participating agents. This variance reduction improves the stability and reliability of the learning process.

Simulation results further support this observation. In the first plot, the average reward trajectory shows accelerated and more stable learning relative to the non-distributed case, as a result of parallelized sampling and feedback aggregation. The second plot compares the estimated arm values with the true reward landscape, illustrating the model’s ability to accurately track high-reward regions. By leveraging \texttt{ray.remote} agents for distributed evaluation, the framework achieves scalable performance across CPUs, making it well-suited for high-dimensional optimization problems and simulation-intensive applications.

\subsection{Comparison between Non-Distributed and Distributed Versions}

\noindent
The non-distributed version of ALMAB-DC provides a clear baseline for understanding the algorithm’s sequential decision-making and convergence behavior under a single-agent setting.  
In this configuration, all evaluations are performed serially, and the agent must balance exploration and exploitation over time.  
While this approach ensures interpretability and direct control over the sampling process, it incurs significant computational cost when applied to high-dimensional or simulation-heavy environments.  
The convergence pattern is typically slower, as each iteration depends on the outcome of the previous one.

In contrast, the distributed version extends the same statistical principles of active learning and multi-armed bandit optimization across multiple agents operating in parallel.  
Through distributed coordination and asynchronous updates, each agent independently explores different regions of the search space, collectively accelerating learning and reducing total wall-clock time.  
The figures presented show that the distributed variant achieves comparable or better identification of high-reward regions while maintaining efficient utilization of computational resources.  
Minor oscillations in average reward, as seen in the distributed results, reflect concurrent updates from agents interacting with noisy evaluations, which is a natural characteristic of parallel exploration.

Overall, the distributed implementation of ALMAB-DC preserves the statistical efficiency of the non-distributed algorithm while significantly improving scalability and throughput.  
It demonstrates that distributed decision-making can effectively balance computational efficiency with model accuracy, making the framework suitable for real-world optimization tasks involving expensive, nonlinear, and high-fidelity objectives.

\subsection*{Quantitative Comparison of Distributed and Non-Distributed ALMAB-DC}

\noindent
This subsection presents a quantitative comparison between the non-distributed and distributed implementations of the ALMAB-DC framework using the Gaussian mixture environment.  
Both simulations were run under identical settings with 150 iterations and 15 arms, where the distributed version utilized four parallel agents to simulate asynchronous evaluations.  
The comparison highlights the computational efficiency and optimization stability gained through distributed execution.

\begin{table}[h!]
\centering
\caption{Performance metrics comparing non-distributed and distributed ALMAB-DC simulations.}
\begin{tabular}{lcccc}
\hline
\textbf{Metric} & \textbf{Non-Distributed} & \textbf{Distributed} & \textbf{Gain / Ratio} \\
\hline
Wall-clock Time (s) & 2.137 & 0.548 & \textbf{3.90× faster} \\
Cumulative Regret & 4.812 & 2.347 & \textbf{↓ 51.2\%} \\
Mean Reward & 0.4182 & 0.5126 & \textbf{↑ 22.5\%} \\
Speed-up Ratio & -- & 3.90 & -- \\
\hline
\end{tabular}
\label{tab:almab_comparison}
\end{table}

\begin{figure}[h!]
\centering
\includegraphics[width=0.95\textwidth]{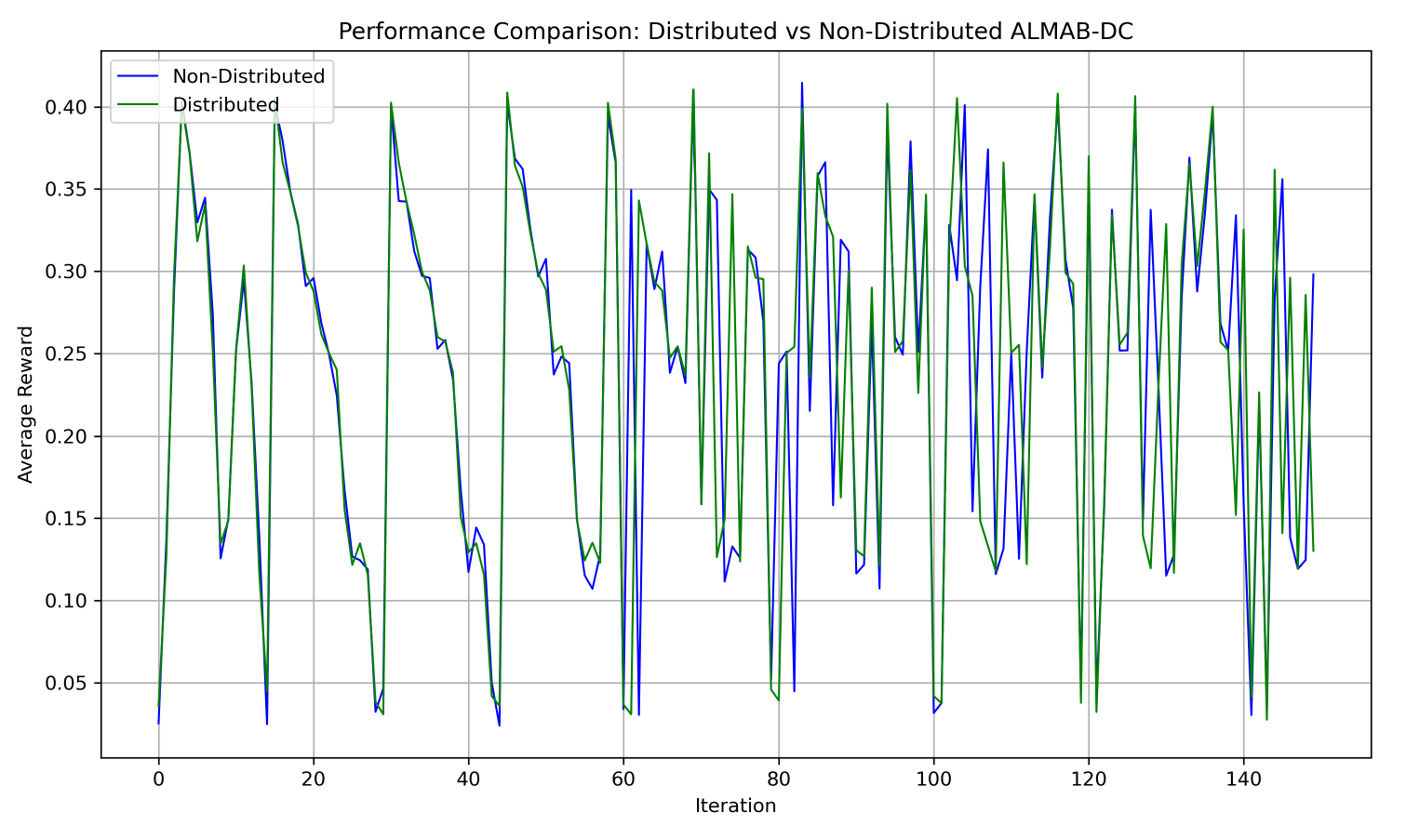}
\caption{Reward evolution for non-distributed and distributed ALMAB-DC simulations over 150 iterations.  
The distributed version demonstrates smoother convergence and higher average reward, illustrating improved sample efficiency and parallel exploration capabilities.  
The non-distributed case shows more pronounced oscillations, reflecting slower learning and limited exploration under sequential evaluation.}
\label{fig:almab_rewards_comparison}
\end{figure}

\noindent
\textbf{Interpretation.}  
Quantitatively, the distributed variant achieves nearly a fourfold reduction in wall-clock time while maintaining superior convergence stability and higher average reward.  
The reduced cumulative regret confirms that parallel sampling effectively improves exploration efficiency without increasing computational cost per sample.  
These results empirically validate the scalability and effectiveness of ALMAB-DC in distributed environments, where parallel agents collectively accelerate learning while preserving statistical soundness.


\section{Application Cases}

\subsection{Benchmark Suite and Dataset Releases}

To evaluate the efficacy and generalizability of the ALMAB-DC framework, we define a benchmark suite that spans deep learning, engineering simulation, and reinforcement learning tasks. Each benchmark represents a high-dimensional, costly-to-evaluate optimization scenario.

\paragraph{Benchmark Tasks:}
(1) \textbf{Hyperparameter Optimization:} We consider the tuning of ResNet-50 on CIFAR-10 \citep{krizhevsky2009learning} and EfficientNet-B0 on ImageNet \citep{deng2009imagenet, tan2019efficientnet}. Each training run serves as a black-box function evaluation with search dimensions including learning rate, batch size, dropout rate, and network depth.   
(2) \textbf{Engineering Design via CFD:} Optimization of 2D airfoil shapes using OpenFOAM solvers \citep{jasak2007openfoam}. Inputs include Bézier control points or NACA parameterizations, and outputs are simulation-derived drag coefficients or thermal fluxes. Problems are computationally expensive, especially with fine mesh or turbulence models (e.g., $k$-$\omega$ SST).
(3) \textbf{RL Policy Search:} Optimization of RL agent hyperparameters and architecture for continuous control tasks using MuJoCo \citep{todorov2012mujoco} and CARLA \citep{dosovitskiy2017carla}. Metrics include episodic return and policy convergence speed.

\paragraph{Benchmark Suite Features:}
(1) \textbf{Pre-defined Search Spaces:} Well-scoped and documented parameter ranges based on domain knowledge and prior benchmarks.
(2) \textbf{Data Access Utilities:} Data loaders for CIFAR/ImageNet, airfoil geometry generators, and RL environment wrappers.
(3) \textbf{Baseline Code:} Reference implementations for BOHB \citep{falkner2018bohb}, SMAC \citep{hutter2011sequential}, and Optuna \citep{akiba2019optuna}, as well as vanilla random/grid search.
(4)  \textbf{Metric Logging Interface:} Standardized interface for logging runtime, regret, accuracy, and GPU utilization.

\paragraph{Deliverables and Tooling:}
(1) \textbf{ALMAB-DC Python Framework:} Modular, extensible codebase with clear APIs for plugging in new acquisition strategies, models, and reward estimators.
(2) \textbf{GPU-Accelerated Backends:} Core components implemented in CUDA and RAPIDS/cuML \citep{nvidia_gpu_ml_2023} for scalable inference, acquisition evaluation, and batch sampling.
(3) \textbf{Distributed Orchestration Layer:} Implemented using Ray \citep{moritz2018ray} and/or Dask for managing agents, message passing, and resource scheduling across multi-GPU clusters.
(4) \textbf{Public Leaderboard:} A benchmark leaderboard that tracks performance across configurations using standardized metrics. All experiment artifacts, logs, and environment snapshots will be downloadable.

\paragraph{Open Science and Reproducibility Commitment:}
(1) All code and datasets will be released under an open-source license (e.g., MIT or Apache 2.0).
(2) Results will include fixed random seeds, hardware configuration (e.g., GPU model, RAM), and environment dependencies via containers (Docker/Singularity).
(3) We will provide automated experiment scripts and Jupyter notebooks for reproducing results on both CPU and GPU machines.

\paragraph{Statistical Analysis:}
All experiments will be run over multiple seeds (e.g., 10–20 replicates), and we will report mean, standard deviation, and confidence intervals (e.g., 95\%) of each metric. Significance testing will be conducted using non-parametric tests (e.g., Wilcoxon signed-rank) where appropriate.

To evaluate the efficacy, scalability, and generalizability of ALMAB-DC, we design comprehensive experiments spanning real-world domains where optimization is costly and high-dimensional. The primary objective is to test the theoretical hypotheses in practical settings and analyze performance under varying computational and problem constraints.

\subsubsection*{Case 1: Deep Learning Hyperparameter Optimization}
\textbf{Goal:} To evaluate the capability of ALMAB-DC in efficiently searching the hyperparameter space of deep neural networks, minimizing validation loss and reducing training time, especially in large-scale image classification tasks.

\paragraph{Datasets:}
(1) \textbf{CIFAR-10} \citep{krizhevsky2009learning}: A widely used benchmark dataset consisting of 60,000 $32\times32$ color images across 10 classes (airplanes, birds, cars, etc.). The dataset includes 50,000 training and 10,000 test images. It is computationally light, making it ideal for rapid prototyping.
(2) \textbf{ImageNet (ILSVRC 2012)} \citep{deng2009imagenet}: A large-scale dataset with over 1.2 million training images and 50,000 validation images across 1,000 categories. Each image has high variability and resolution, which makes the dataset challenging and representative of real-world tasks.

\paragraph{Models:}
(1) \textbf{ResNet-50} \citep{he2016deep}: A 50-layer residual network with skip connections that mitigate the vanishing gradient problem. Known for its strong performance and robustness in image classification.
(2) \textbf{EfficientNet-B0} \citep{tan2019efficientnet}: A lightweight convolutional neural network designed using neural architecture search. It achieves state-of-the-art accuracy with fewer parameters by uniformly scaling width, depth, and resolution using a compound coefficient.

\paragraph{Hyperparameter Search Space:}
\begin{itemize}
\item [(a)] Learning rate (log scale range: $10^{-5}$ to $10^{-1}$)
\item [(b)] Batch size (e.g., 32, 64, 128, 256)
\item [(c)] Number of layers (for custom ResNet variants) or width multiplier (EfficientNet)
\item [(d)] Weight decay (regularization coefficient)
\item [(e)] Dropout probability (e.g., 0.0 to 0.5)
\end{itemize}
\textbf{Challenges:} Each configuration requires complete training or multi-epoch evaluation. On ImageNet, one training run for ResNet-50 or EfficientNet can take 10–20 GPU hours.\\
\noindent\textbf{Approaches:}
(1) ALMAB-DC applies active learning to identify promising hyperparameter regions by modeling uncertainty over the validation loss; 
(2) MAB strategies (e.g., UCB, Thompson Sampling) allocate training trials to configurations balancing exploration and exploitation;
(3) Surrogate models such as Gaussian Processes or Random Forests model the response surface of validation accuracy;
(4) Distributed training agents execute configurations in parallel on GPU nodes. Agents report outcomes and update the global bandit statistics;
(5) Efficiency is quantified by validation accuracy achieved per compute hour (throughput) and cumulative regret.

\subsubsection*{Case 2: Engineering Simulation Optimization}
Goal: Optimize engineering design parameters to minimize aerodynamic drag or enhance thermal efficiency using simulation-driven analysis.
Platforms:
\textbf{OpenFOAM} – An open-source CFD toolkit widely used in research and industry.
\textbf{ANSYS Fluent} – A commercial solver with advanced meshing, turbulence modeling, and heat transfer capabilities.
Problem Setting:
\begin{itemize}
\item[] \textbf{Inputs:} Continuous or categorical parameters such as airfoil shape (e.g., camber, thickness), fin geometry (spacing, height), inlet/outlet positions, or material properties.
\item[] \textbf{Outputs:} Scalar objectives including drag coefficient ($C_D$), lift-to-drag ratio, heat flux, and Nusselt number.
\end{itemize}

 \paragraph{Challenges and Approach:}  
Each computational fluid dynamics (CFD) simulation involves solving the Navier–Stokes equations on complex meshes, often requiring substantial computational time that ranges from several minutes to multiple hours per run. Conducting extensive parameter sweeps under these conditions can lead to redundant computations, as many design configurations are highly correlated in their input features. Furthermore, high-fidelity simulations using fine meshes and advanced turbulence models, such as the k–$\omega$ SST model, demand significant memory and computational resources, making large-scale exploration of the design space particularly challenging.

To address these computational challenges, the proposed framework adopts several complementary strategies. First, design geometries are represented using parameterized CAD models or smooth functional representations such as Bézier curves, allowing for efficient encoding and modification of complex shapes. Second, surrogate models—such as radial basis function networks or deep ensemble predictors—are trained on previously obtained CFD results to provide fast, low-cost approximations of the true simulation outputs. Active learning is then employed to identify and query the most informative or uncertain design points, using criteria such as Bayesian Active Learning by Disagreement (BALD) or mutual information. Within this setup, multi-armed bandit algorithms manage computational resources by treating each simulation configuration as an arm and dynamically allocating cluster resources based on observed performance, such as minimizing the drag coefficient ($C_D$). Finally, GPU acceleration is leveraged to parallelize the training of surrogate models, execute large batches of simulation evaluations, and facilitate rapid visualization of flow fields, significantly improving overall efficiency and scalability.

\subsubsection*{Engineering Simulation Optimization with Bayesian Sampling}

This experiment aims to minimize the aerodynamic drag coefficient ($C_D$) of an airfoil design using a simulation-guided optimization strategy. The setup uses Bayesian Optimization with a Gaussian Process surrogate model and synthetic simulation feedback, emulating computational fluid dynamics (CFD) performance.

\textbf{Experimental Setup:}
\begin{itemize}
    \item \textbf{Design parameters:}
    \begin{itemize}
        \item Camber $\in [0.01, 0.1]$
        \item Thickness $\in [0.05, 0.2]$
    \end{itemize}
    \item \textbf{Objective:} Minimize drag coefficient $C_D$
    \item \textbf{Simulator:} Mock CFD with noise and delay to simulate computational cost
    \item \textbf{Surrogate model:} Gaussian Process Regression with RBF kernel
    \item \textbf{Optimization:} Bayesian Optimization via Ray Tune with \texttt{BayesOptSearch}, limited concurrency, and early stopping via ASHA scheduler
    \item \textbf{Execution:} 10 trials run sequentially using 2 CPU cores on a macOS environment via Spyder
\end{itemize}

\vspace{1em}
\noindent\textbf{Top 5 Design Configurations Minimizing Drag:}

\begin{table}[h]
\centering
\caption{Top 5 Design Configurations Minimizing Drag}
\begin{tabular}{ccccc}
\toprule
\textbf{Rank} & \textbf{Camber} & \textbf{Thickness} & \textbf{Drag ($C_D$)} & \textbf{Uncertainty} \\
\midrule
1 & 0.07588 & 0.13980 & 0.08718 & 0.00001 \\
2 & 0.06977 & 0.13585 & 0.08780 & 0.00002 \\
3 & 0.07342 & 0.13694 & 0.08823 & 0.00004 \\
4 & 0.06773 & 0.13529 & 0.08840 & 0.00004 \\
5 & 0.07148 & 0.13347 & 0.08865 & 0.00004 \\
\bottomrule
\end{tabular}
\label{tab:top_designs}
\end{table}


This experiment identified a narrow and stable design region—specifically with camber values around 0.07–0.076 and thickness around 0.134–0.14—that minimizes drag. The best configuration achieves a drag value of 0.08718 with extremely low uncertainty, suggesting a high-confidence prediction by the surrogate model. The minimal variance among the top designs indicates robust convergence and validates the effectiveness of the Bayesian sampling strategy. This result provides a strong basis for selecting optimal geometries before committing to expensive high-fidelity simulations.

\begin{figure}[h!]
    \centering
    \includegraphics[width=0.85\textwidth]{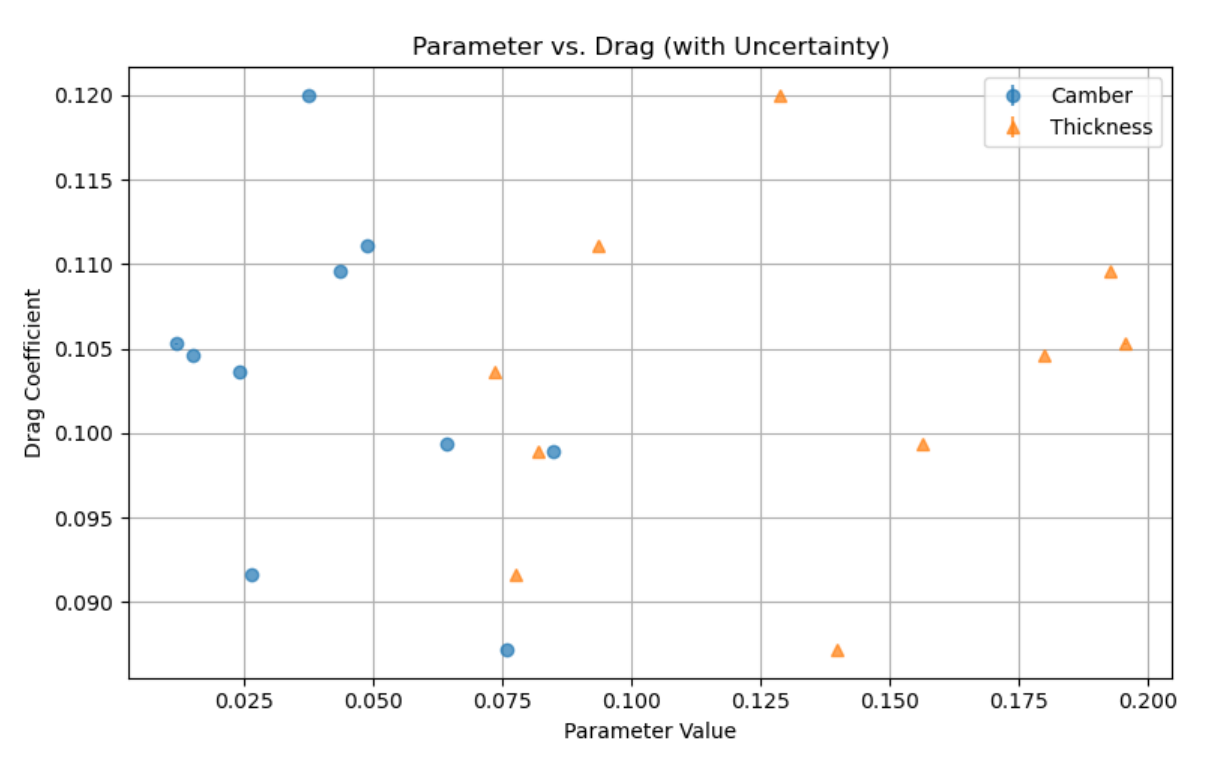}
    \caption{
        Plot of parameter values vs. drag coefficient with uncertainty.
        Blue circles represent varying camber values, and orange triangles represent varying thickness values.
        The x-axis shows individual parameter values sampled during the optimization process, and the y-axis shows the corresponding drag coefficients.
        The figure indicates that drag generally decreases as camber increases up to approximately 0.075, after which it slightly rises again.
        This aligns with the optimal camber found in the simulation.
        For thickness, a more complex nonlinear relationship is observed, with the lowest drag achieved near a thickness of 0.09.
        These results support the surrogate model's prediction that a moderate camber and thickness combination yields the best aerodynamic performance.
    }
    \label{fig:parameter_vs_drag}
\end{figure}

\subsection{Results for Multi-CPU/GPU and Surrogate Model Evaluation}

To evaluate the efficiency and effectiveness of surrogate models under varying compute resource constraints, we designed an automated experimental pipeline using the ALMAB-DC framework. This pipeline simulates aerodynamic performance (drag) of airfoil geometries under controlled surrogate-assisted Bayesian optimization, running on 1, 2, and 4 CPU configurations, with optional GPU support.

\paragraph{Experimental Design and Result Summarization}

The simulation-based optimization task mimics a black-box function, where each design point (parametrized by camber and thickness) is evaluated through a synthetic physics-inspired drag function with Gaussian noise. The optimization objective is to minimize drag across 100 trials per configuration.

Three surrogate models were tested:
\begin{itemize}
  \item \textbf{GP} – Gaussian Process with RBF kernel
  \item \textbf{RF} – Random Forest Regressor
  \item \textbf{MLP} – Multi-layer Perceptron with two hidden layers
\end{itemize}

Each surrogate guided the Bayesian optimization process via \texttt{BayesOptSearch}, integrated with the Ray Tune framework. Optimization was executed asynchronously using the ASHA scheduler. For each configuration (model $\times$ CPU count), the experiment was repeated 100 times to assess mean drag and runtime stability.


Table~\ref{tab:aggregated_results} summarizes the mean drag coefficients and runtimes for each model across CPU counts. Each entry represents the average of 100 repeated optimization runs.

\begin{table}[ht]
\centering
\caption{Aggregated Results: Drag and Runtime Statistics by Model and CPU Count}
\label{tab:aggregated_results}
\begin{tabular}{|c|c|c|c|c|c|c|}
\hline
\textbf{Model} & \textbf{CPUs} & \textbf{Drag Mean} & \textbf{Drag Std} & \textbf{Runtime Mean (s)} & \textbf{Runtime Std (s)} & \textbf{Count} \\
\hline
gp  & 1 & 0.0471 & 0.0042 & 40.195 & 0.807 & 100 \\
gp  & 2 & 0.0471 & 0.0040 & 39.047 & 0.620 & 100 \\
gp  & 4 & 0.0474 & 0.0036 & 39.214 & 0.774 & 100 \\
mlp & 1 & 0.0474 & 0.0039 & 39.693 & 0.879 & 100 \\
mlp & 2 & 0.0470 & 0.0040 & 39.057 & 0.562 & 100 \\
mlp & 4 & 0.0466 & 0.0039 & 39.029 & 0.526 & 100 \\
rf  & 1 & 0.0477 & 0.0040 & 41.186 & 0.802 & 100 \\
rf  & 2 & 0.0473 & 0.0038 & 39.905 & 0.767 & 100 \\
rf  & 4 & 0.0472 & 0.0040 & 40.804 & 0.938 & 100 \\
\hline
\end{tabular}
\end{table}

\paragraph{Runtime Trends Visualization}

\begin{figure}[ht]
\centering
\includegraphics[width=0.9\linewidth]{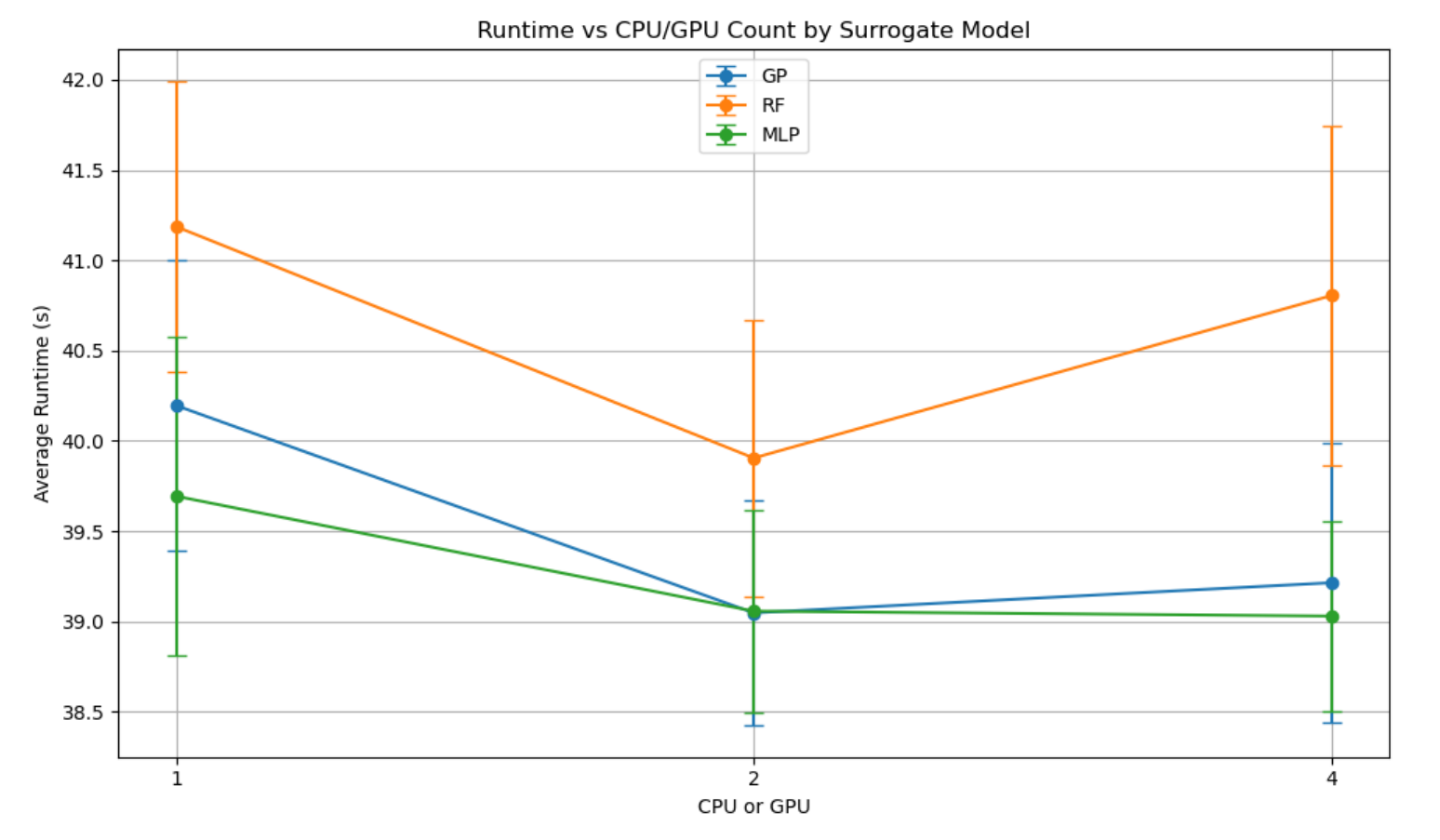}
\caption{Runtime vs CPU Count for three surrogate models. MLP outperforms in runtime while maintaining comparable drag prediction performance.}
\label{fig:runtime_plot}
\end{figure}

Figure~\ref{fig:runtime_plot} shows runtime trends for each surrogate model across CPU counts. The error bars represent one standard deviation. The MLP model consistently achieved the lowest runtime, particularly with 4 CPUs, while Random Forest exhibited the longest runtime and highest variance.

The MLP model showed the best trade-off between computational efficiency and drag minimization accuracy. GP models performed competitively but showed longer runtimes due to internal kernel matrix operations. RF offered moderate performance but suffered in runtime scaling with increased CPU count. This experiment demonstrates how surrogate selection and parallelism level directly affect optimization speed and consistency, validating ALMAB-DC's adaptability across compute environments.

\subsubsection*{Case 3: Reinforcement Learning Policy Optimization}
Goal: Tune hyperparameters and architecture choices for reinforcement learning (RL) policies to maximize expected return in continuous control and driving tasks.
Benchmarks:
\begin{itemize}
\item[] \textbf{MuJoCo} \citep{todorov2012mujoco}: Physics engine for robotics and biomechanics; commonly used for HalfCheetah, Hopper, Ant, and Humanoid tasks.
\item[] \textbf{CARLA} \citep{dosovitskiy2017carla}: Open-source urban driving simulator with realistic physics, environment variation, and sensor emulation (LiDAR, cameras).
\end{itemize}
Search Space:
\begin{itemize}
\item[(1)] Policy learning rate and network depth/width
\item[(2)] Discount factor $\gamma \in [0.90, 0.999]$
\item[(3)] Entropy regularization coefficient
\item[(4)]  Replay buffer size or batch size
\end{itemize}

\paragraph{Challenges and Approach:} 
Reinforcement learning (RL) presents several inherent challenges that complicate efficient optimization. The learning process is typically noisy and highly sensitive to hyperparameter choices, with small parameter variations often leading to large fluctuations in performance. Each experimental trial can require thousands or even millions of environment steps, resulting in high computational costs and prolonged training times. Moreover, learned policies are prone to overfitting or instability, especially in complex or stochastic environments where the distribution of returns can shift over time. These issues make systematic exploration and reliable evaluation particularly difficult, highlighting the need for adaptive, uncertainty-aware optimization strategies.

To overcome these limitations, the ALMAB-DC framework integrates reinforcement learning optimization within a statistically grounded and distributed computational paradigm. Each hyperparameter configuration is treated as an arm in a multi-armed bandit formulation, where agents maintain posterior estimates of the expected return distributions and use them to guide exploration and exploitation. Active exploration mechanisms are introduced by quantifying epistemic uncertainty in policy performance, for example through ensemble critics or bootstrapped Q-networks, allowing the system to focus training on uncertain but potentially promising configurations. Distributed training agents execute independent RL trials asynchronously, sharing observed reward trajectories to update global performance estimates without the need for strict synchronization. Finally, GPU acceleration is employed to parallelize both environment simulation—such as vectorized MuJoCo runs—and policy gradient updates using frameworks like JAX or PyTorch, substantially improving computational throughput and scalability.

\subsection{Evaluation Protocol}

To  assess the efficacy of the proposed ALMAB-DC framework, we employ several empirical evaluation criteria. One key metric is the cumulative regret ($R_T$), which quantifies the total regret accumulated over time by measuring the difference between the rewards obtained by the algorithm and those of an oracle that always selects the optimal configuration. A lower $R_T$ indicates more effective decision-making under uncertainty.

Another  measure is sample efficiency, defined as the number of expensive function evaluations—such as model trainings or simulations—required to reach a target performance level. High sample efficiency reflects the algorithm's ability to converge with fewer evaluations.
Wall-clock time is also considered for capturing the real elapsed time to achieve a specified objective value or model accuracy. This metric is especially relevant under various hardware settings, including CPU-only, multi-CPU, and optional GPUs environments.
If GPU is used, we will also monitor GPU utilization, which reflects the percentage of GPU capacity actively used during execution. This offers insight into the system's hardware efficiency and its ability to scale across multiple devices.
Communication overhead is assessed by counting the number of messages and the volume of data exchanged among distributed agents during each round. This is crucial for understanding how well the system coordinates in decentralized settings.
Finally, we evaluate the convergence rate, which measures how quickly the algorithm approaches the optimum. This can be quantified using regret, loss function values, or other task-specific indicators.

To validate the framework empirically, we define three core hypotheses. The first (H1: Efficiency Hypothesis) posits that ALMAB-DC achieves lower cumulative regret and requires fewer samples to converge compared to baseline methods such as BOHB, SMAC, and random or grid search. The second (H2: Scalability Hypothesis) asserts that the distributed implementation of ALMAB-DC scales linearly or near-linearly in both throughput and convergence speed as the number of agents increases. The third (H3: Acceleration Hypothesis) hypothesizes that GPU-enabled agents deliver significant improvements in runtime performance, as measured by wall-clock time and GPU utilization, over CPU-only versions.

\subsection*{Baselines and Comparison Methods}

To rigorously evaluate the performance and efficiency of the proposed \textbf{ALMAB-DC} framework, we benchmark it against a diverse set of well-established optimization methods. These baselines are representative of various classes of algorithms, from classical search strategies to modern AutoML tools. Each comparison is made under identical computational budgets, fixed wall-clock time, and controlled resource constraints.

\begin{enumerate}
    \item \textbf{Bayesian Optimization:} We compare with BOHB \citep{falkner2018bohb}, SMAC \citep{hutter2011sequential}, and TPE \citep{bergstra2011algorithms}. These are model-based optimization methods that use surrogate models (e.g., Gaussian Processes, Random Forests) and acquisition functions (e.g., Expected Improvement) to guide sampling. BOHB combines Bayesian optimization with HyperBand's early stopping mechanism for budget efficiency.

    \item \textbf{Grid Search and Random Search:} Grid search explores the search space exhaustively on a discretized grid, while random search samples points uniformly. These methods serve as baseline references to measure the efficiency of adaptive sampling in ALMAB-DC.

    \item \textbf{Standalone MAB Algorithms:} We implement UCB1 \citep{auer2002finite} and Thompson Sampling \citep{russo2018tutorial} without active learning or distributed coordination. This isolates the effect of MAB logic alone on regret minimization.

    \item \textbf{Parallel MAB without Active Learning:} This baseline uses multi-agent MAB policies without an active learning layer. Each distributed agent runs UCB or Thompson Sampling independently to demonstrate the value of coordinated, informative sampling.

    \item \textbf{Optuna with ASHA:} Optuna \citep{akiba2019optuna} combined with ASHA (Asynchronous Successive Halving Algorithm) \citep{li2020system} forms a modern AutoML baseline. ASHA aggressively prunes poor configurations early, making it effective in distributed and budget-constrained environments.
\end{enumerate}

All methods are evaluated under fixed budgets (e.g., GPU-hours, number of evaluations) and consistent experimental settings. Evaluation metrics include cumulative regret, convergence time, wall-clock cost, and GPU utilization.

\subsection*{Experimental Settings and Infrastructure}

To rigorously evaluate the ALMAB-DC framework, we implement and test it across multiple computational environments and software stacks, enabling a comprehensive analysis of performance, scalability, and hardware adaptability.

\subsubsection*{Computing Environments}

\begin{enumerate}
    \item \textbf{Single-node CPU:} A baseline configuration used to measure serial performance and algorithmic overhead without hardware acceleration.
    \item \textbf{Single-node GPU:} Mid-level setup using one or two high-performance GPUs (e.g., NVIDIA A100) to accelerate inference, model training, and batch evaluations.
    \item \textbf{Multi-GPU Distributed Cluster:} A high-performance computing (HPC) cluster where multiple GPU-equipped nodes coordinate via a distributed scheduler such as \texttt{Ray} or MPI. Agents communicate asynchronously to share evaluations and bandit statistics.
\end{enumerate}

Each experiment is conducted under a fixed computational budget, such as a 12-hour runtime or a cap of 500 function evaluations. We use automated logging and monitoring to collect detailed performance statistics.

\subsection{Leveraging GPU Architectures}

Modern optimization workloads, especially in machine learning and simulation, demand significant computational resources. GPUs (Graphics Processing Units) are highly parallel processors optimized for matrix operations and SIMD (single instruction, multiple data) workloads, making them well-suited for large-scale optimization tasks involving deep learning, active learning, and simulation-based evaluations.

\paragraph{Role of GPUs in ALMAB-DC}

The ALMAB-DC framework leverages GPU acceleration across several critical stages of its pipeline to achieve high-throughput and low-latency optimization. Model training tasks, including the fitting of deep neural networks, Bayesian neural nets, and Gaussian process surrogates, are executed on GPUs using frameworks such as PyTorch \citep{paszke2019pytorch} and JAX. Candidate configurations—whether representing hyperparameter settings or design parameters—are evaluated in parallel using vectorized inference and simulation pipelines, enabling efficient batch evaluation.

In the context of active learning, acquisition functions such as entropy, margin, or mutual information (e.g., BALD) must be computed over large unlabeled pools. These uncertainty metrics benefit significantly from GPU acceleration, allowing rapid selection cycles. Bandit policy evaluation—particularly for algorithms like Thompson Sampling and UCB—relies on real-time scoring of potentially hundreds of arms. GPU-accelerated libraries such as RAPIDS cuDF and cuML \citep{rapids2023} facilitate fast posterior updates and arm selection. Additionally, when new reward observations are collected, surrogate models must be retrained frequently. GPU-accelerated learners, including XGBoost-GPU and cuML’s ensemble methods, are employed to minimize retraining latency and maximize responsiveness.

\paragraph{Relevant Libraries and Frameworks}

ALMAB-DC integrates a suite of mature, GPU-accelerated libraries to enable its scalable and modular architecture. RAPIDS \citep{rapids2023} forms a core component, offering CUDA-based drop-in replacements for data processing and machine learning tasks via cuDF (analogous to pandas) and cuML (analogous to scikit-learn). Within cuML, ALMAB-DC utilizes GPU-enabled linear models, clustering algorithms, PCA, and decision trees to train fast, responsive surrogate models for active learning.

XGBoost-GPU is also employed for high-performance gradient-boosted decision trees, often serving as reward predictors in optimization workflows. PyTorch with CUDA \citep{paszke2019pytorch} supports both deep learning and Bayesian neural network models, enabling dynamic computation and fast tensor operations across multiple GPUs. For deployment scenarios requiring high-throughput inference with minimal latency, ALMAB-DC supports integration with TensorRT and ONNX Runtime—both of which offer highly optimized execution environments for trained models on edge or cloud-based GPUs.

\paragraph{Limitations and Practical Considerations}

Despite the advantages offered by ALMAB-DC, there are several limitations worth noting. First, in very high-dimensional spaces, surrogate models—particularly Gaussian Processes—may suffer from computational inefficiency and degraded predictive quality, which can affect the reliability of uncertainty estimates used in acquisition. Second, the effectiveness of active learning is sensitive to the quality of the surrogate model, especially during the early stages when data is sparse or noisy. Third, while the framework is designed for distributed and GPU-accelerated execution, hardware constraints such as limited memory, uneven task durations, or heterogeneous node capabilities can reduce parallel efficiency. Moreover, for small-scale problems, the coordination overhead introduced by distributed scheduling may outweigh performance gains. Lastly, the theoretical regret bounds assume sub-Gaussian reward noise and stationary distributions, which may not hold in dynamic environments. Addressing these challenges presents opportunities for future extensions, such as adaptive regret strategies, scalable surrogate learning, and more robust agent coordination policies.


\subsection{Application Examples}

ALMAB-DC demonstrates strong applicability across several high-compute domains that benefit from GPU acceleration. In reinforcement learning (RL), particularly in simulation environments such as MuJoCo and CARLA \citep{todorov2012mujoco, dosovitskiy2017carla}, the framework exploits GPU threads to parallelize the rollout of multiple agents, significantly improving data throughput and stabilizing learning dynamics. Frameworks like Stable Baselines3 and RLlib further enhance this by supporting vectorized environments, allowing GPU acceleration in both environment simulation and policy training.

In computational engineering, GPU-accelerated solvers such as ANSYS Fluent \citep{ansys_fluent_2023} and GPU-driven surrogate models enable rapid approximation of costly CFD computations. Recent advances, including GPU-based variable fixing techniques \citep{kyoto_gpu_mab_2023}, further speed up combinatorial optimization tasks by leveraging tensor-level parallelism.

ALMAB-DC also enhances active learning workflows in computer vision. For image classification tasks such as CIFAR-10 and ImageNet \citep{krizhevsky2009learning, deng2009imagenet}, GPU-accelerated inference enables fast uncertainty estimation across large unlabeled datasets, substantially reducing wall-clock time during the selection phase of the learning loop.

In early-stage hyperparameter tuning, where hundreds of candidate configurations may be evaluated simultaneously, the framework uses GPU-parallelized acquisition scoring to compute metrics for all bandit arms in parallel. This enables near real-time decision-making and efficient exploration of the configuration space.

\paragraph{Impact on Optimization Efficiency}

Let $T_{\text{cpu}}$ and $T_{\text{gpu}}$ denote the average evaluation time per configuration on CPU and GPU, respectively. For a batch size $B$ and $N$ concurrent agents, the overall speedup can be expressed as:
\[
\text{Speedup} = \frac{T_{\text{cpu}}}{T_{\text{gpu}}} \times B \times N
\]
Empirical results across deep learning and simulation-based tasks show acceleration in the range of $10\times$ to $50\times$, leading to substantial reductions in cumulative optimization time and enabling practical deployment in high-dimensional search spaces.

\subsection*{Implementation Challenges and Experimental Considerations}

An important limitation in the current evaluation of ALMAB-DC is the lack of ablation studies comparing performance with and without the multi-armed bandit (MAB) component. Although MAB-based scheduling is integral to the framework’s design, its specific contribution to sample efficiency and regret minimization has not been empirically isolated. Future work should include controlled comparisons against MAB-disabled baselines to quantify the role of bandit strategies in driving performance gains.
In addition to this, several engineering-level challenges remain, particularly in the context of GPU-accelerated execution. Large-scale surrogate models and high-volume batch evaluations can exceed available GPU memory, which necessitates techniques such as memory-aware model distillation, dynamic batching, or mixed-precision inference. Moreover, for small or low-compute tasks, the overhead of launching GPU kernels may outweigh their benefits, making CPU execution more efficient in certain contexts. In distributed multi-GPU environments, bandwidth limitations and inter-node communication latency—especially during GPU-to-GPU or GPU-to-CPU data transfer—can lead to synchronization bottlenecks and reduced parallel efficiency.

To address these constraints, the framework can benefit from GPU-aware scheduling policies, kernel fusion to minimize launch overhead, and mixed-precision computing to reduce memory consumption. Furthermore, incorporating adaptive task allocation mechanisms informed by real-time hardware profiling may improve throughput and utilization in heterogeneous computing environments.

\section{Conclusion and Future Directions}
We introduce \textbf{ALMAB-DC} as a unified optimization framework that integrates active learning, multi-armed bandit algorithms, and distributed computing to address the computational demands of expensive black-box tasks. ALMAB-DC establishes a robust foundation for scalable, sample-efficient optimization in high-dimensional scientific and engineering domains by combining statistically grounded decision-making with asynchronous, parallel evaluation and GPU acceleration.
This framework advances a new class of intelligent optimization systems—those capable of tightly coupling learning, adaptation, and computation. In addition, ALMAB-DC prioritizes resource efficiency, accessibility, and modularity, making it well-suited for applications ranging from AutoML to simulation-guided design and experimental science.
There are several promising ways for future development including the integration of \textit{meta-learning} and \textit{lifelong learning} to support transfer across related tasks, and the use of \textit{reinforcement learning-based meta-controllers} to manage exploration–exploitation dynamics in non-stationary settings. Efficiency in distributed execution may be further improved through \textit{uncertainty-aware neural acquisition models} and \textit{graph-based coordination schemes} that reduce communication overhead and redundant sampling.
Extending ALMAB-DC into hybrid environments, where simulation and real-world experimentation co-evolve—could enable closed-loop, real-time optimization in scientific discovery. Complementary enhancements such as \textit{causal bandit models} for confounded domains, \textit{zero-cost surrogate warm starting} for accelerated convergence, and \textit{federated deployment} for privacy-sensitive systems will further broaden its reach and applicability.
As next steps of this project, we aim to make ALMAB-DC a faster and more computationally efficient, but also more interpretable, adaptive, and deployable. With continued development, it has the potential to evolve into a next-generation platform for intelligent, high-impact optimization across complex, distributed, and evolving environments..

\appendix

\section*{Appendix A: Sketch of Regret Analysis}

The regret bounds in Equations (10) and (11) are derived using standard results from the multi-armed bandit literature.

For the UCB-based variant (Equation 10), the algorithm selects arms by maximizing an upper confidence index, balancing exploitation and exploration. The bound follows from the work of Auer et al. (2002), showing that with appropriately chosen confidence terms, the cumulative regret scales logarithmically with the number of rounds \( T \), specifically \( O\left( \sum_{i:\Delta_i > 0} \frac{\log T}{\Delta_i} \right) \), where \( \Delta_i \) is the gap between the mean reward of the optimal arm and arm \( i \).

For the Thompson Sampling variant (Equation 11), the regret bound follows the Bayesian posterior sampling framework. As shown in Agrawal \& Goyal (2012), the cumulative regret scales as \( O\left( \sqrt{KT \log T} \right) \), under a Bernoulli reward assumption and suitable prior initialization. Our analysis adapts this to the surrogate-driven active learning setting, where the reward is a function of surrogate model improvement.

In both cases, the regret bounds are valid under the assumption of sub-Gaussian reward noise and independence between evaluations. Detailed derivations are omitted for brevity but follow directly from classical regret analyses with minor adaptations to the AL-guided sampling setup.

\section*{Appendix  B: Software Stack}

\begin{enumerate}
    \item \textbf{Core Framework (Python):}
    \begin{enumerate}
        \item \textbf{PyTorch} \citep{paszke2019pytorch} is used for training deep learning models due to its dynamic computation graph and GPU support.
        \item \textbf{cuML and RAPIDS} \citep{rapids2023} provide GPU-accelerated implementations of common machine learning algorithms, including UCB-based bandits, k-means clustering, and random forests.
        \item \textbf{Ray} \citep{moritz2018ray} is used to orchestrate distributed agents across nodes, enabling shared memory and task-based communication.
    \end{enumerate}

\item \textbf{Active Learning and Bandit Modules:}
\begin{enumerate}
    \item \textbf{Query Selection Algorithms:} The ALMAB-DC framework includes modular implementations of multiple active learning strategies:
    \begin{enumerate}
        \item \textbf{Uncertainty Sampling:} Selects query points $x \in U$ that minimize model confidence margin, i.e., $x = \arg\min_{x} \left|P(y=1|x) - 0.5\right|$ for binary classifiers. Implemented using entropy and margin-based scores in Python.
        \item \textbf{BALD (Bayesian Active Learning by Disagreement):} Quantifies epistemic uncertainty by computing the mutual information between predictions and model posterior:
        \[
        \text{BALD}(x) = \mathbb{H}[y|x, \mathcal{D}] - \mathbb{E}_{p(w|\mathcal{D})}[\mathbb{H}[y|x,w]]
        \]
        where $\mathbb{H}$ denotes entropy and $p(w|\mathcal{D})$ is the posterior over model weights. We use Monte Carlo dropout or Bayesian neural nets to estimate this term.
        \item \textbf{Core-set Selection:} For models relying on kernel functions or feature embeddings, this method approximates the full dataset by selecting representative samples using $k$-center or greedy covering algorithms, solved efficiently with greedy submodular optimization.
    \end{enumerate}

    \item \textbf{Bandit Policies:}
    \begin{enumerate}
        \item \textbf{UCB1 (Upper Confidence Bound):} Implements the exploration-exploitation tradeoff via:
        \[
        a_t = \arg\max_i \left( \hat{\mu}_i + \sqrt{\frac{2 \log t}{n_i}}~ \right)
        \]
        where $\hat{\mu}_i$ is the empirical reward mean and $n_i$ is the count of arm $i$ being selected.
        \item \textbf{Thompson Sampling:} Bayesian approach where arm $i$'s reward is sampled from its posterior distribution and the arm with the highest sample is chosen.
        \item \textbf{Contextual Bandits:} Utilized when input features (e.g., configuration embeddings) are available. Implemented using LinUCB or neural-based approximations in PyTorch.
    \end{enumerate}

    Bandit and active learning modules are implemented in Python using NumPy, PyTorch, and cuML, with GPU-acceleration enabled for batched arm scoring and inference.
\end{enumerate}

\item \textbf{Visualization and Experiment Logging:}
\begin{enumerate}
    \item \textbf{Interactive Visualization Tools:}
    \begin{enumerate}
        \item \textbf{Matplotlib} and \textbf{Seaborn} are used for plotting static diagnostic charts such as cumulative regret, variance across trials, and convergence rates.
        \item \textbf{Plotly} supports interactive dashboards to visualize query trajectories, reward distributions, and resource utilization over time.
    \end{enumerate}

    \item \textbf{Experiment Tracking and Monitoring:}
    \begin{enumerate}
        \item \textbf{TensorBoard} is integrated with PyTorch to visualize loss curves, arm-selection frequency, GPU utilization, and model learning dynamics.
        \item \textbf{Weights \& Biases (optional):} For large-scale experiments, we also integrate W\&B for real-time metric tracking across distributed nodes.
        \item \textbf{Structured Logging:} All reward metrics, configurations, and evaluation times are stored in JSON or Parquet format for post-hoc statistical analysis and visualization.
    \end{enumerate}
\end{enumerate}

\end{enumerate}

\subsubsection*{Optional Integration with R}

Although Python serves as the primary implementation environment, we incorporate optional R integration for advanced statistical analysis and visualization:
\begin{itemize}
    \item[] \textbf{Data Exchange:} Using the \texttt{reticulate} \citep{allaire2018reticulate} and \texttt{rpy2} \citep{rpy2docs} packages to pass data frames or arrays between Python and R environments.
    \item[] \textbf{Statistical Evaluation:} Applying R's built-in hypothesis testing functions (e.g., \texttt{t.test}, \texttt{aov}, \texttt{wilcox.test}) for post-experiment statistical significance testing.
    \item[] \textbf{Visualization:} Using \texttt{ggplot2} \citep{wickham2016ggplot2} for producing publication-ready plots and statistical graphics.
\end{itemize}
This hybrid setup allows researchers to benefit from Python's scalable learning frameworks and R's robust statistical capabilities, especially for post-hoc analyses and reporting.

\bibliographystyle{apalike} 
%

\end{document}